\documentclass[prd,twocolumn,showpacs,preprintnumbers]{revtex4}
\usepackage{amsmath,amsfonts,amssymb,bm}
\usepackage{graphicx} % do wczytywania rysunkow
\usepackage[hang,normalsize,bf]{subfigure} % rysuje kilka rysunkow
\begin{document}

\title{\Huge \bf Photon-jet correlations  \\
in $pp$ and $p \bar p$ collisions \\}

\author{T. Pietrycki}
\affiliation{Institute of Nuclear Physics\\
PL-31-342 Cracow, Poland\\}

\author{A. Szczurek}
\affiliation{Institute of Nuclear Physics\\
PL-31-342 Cracow, Poland\\}
\affiliation{University of Rzesz\'ow\\
PL-35-959 Rzesz\'ow, Poland\\}

\date{\today}

%===============
\begin{abstract}
We compare results of the $k_t$-factorization approach and the
next-to-leading order
collinear-factorization approach for photon-jet correlations in $pp$ and
$p \bar p$ collisions at RHIC and Tevatron energies. We discuss correlations 
in the azimuthal angle as well as in the two-dimensional space of
transverse momentum of photon and jet. Different unintegrated parton
distributions (UPDF)
are included in the $k_t$-factorization approach. The results depend on
UPDFs used. The standard collinear approach gives cross section
comparable to the $k_t$-factorization approach. 
For correlations of the photon and any jet the NLO contributions
dominate at relatively small azimuthal angles as well as for asymmetric
transverse momenta.
For correlations of the photon with the leading jet (the one having the
biggest transverse momentum) the NLO approach gives zero
contribution at $\phi_{-} < \pi/2$ which opens a possibility to study
higher-order terms and/or UPDFs in this region.
\end{abstract}
%=============


\maketitle

%=============================
\section{Introduction}
%=============================

The jet-jet correlations are interesting probe of QCD dynamics
\cite{D0_dijets}. Recent studies of hadron-hadron correlations at
RHIC \cite{RHIC_hadron_hadron} open a new possibility to study the
dynamics of jet and particle production. The hadron-hadron correlations
involve both jet-jet correlations as well as complicated jet structure.
Recently a preliminary data on photon-hadron azimuthal correlation
in nuclear collisions were also presented \cite{RHIC_photon_hadron}.
In principle, such correlations should be easier for theoretical
description as here only one jet enters, at least in leading order pQCD.
On the experimental side, such measurements are more difficult
due to much reduced statistics as compared to the dijet studies.

Up to now no theoretical calculation for photon-jet were presented in
the literature, even for elementary collisions.
In leading-order collinear-factorization approach
the photon and the associated jet are produced back-to-back.
If transverse momenta of partons entering the hard process are
included, the transverse momenta of the photon and the jet are no
longer balanced and finite (non-zero) correlations in a broad range of
relative azimuthal angle and/or in lengths of transverse momenta of 
the photon and the jet are obtained. The finite correlations can be also
obtained in higher-order collinear-factorization approach
\cite{Berends}.
According to our knowledge no detailed studies for present accelerators
were presented in the literature.

In contrast to the coincidence studies the inclusive distributions
of photons were studied carefully in pQCD up to the next-to-leading order
\cite{Aurenche87}.
Similar studies were performed recently also in the $k_t$-factorization
approach \cite{LZ_photon,PS06_photon}. A rather good description of
direct-photon inclusive cross sections can be obtained in both approaches. 
The $k_t$-factorization approach offers a relatively easy method to
calculate the photon-jet correlations \cite{PS06_photon}.

The $k_t$-factorization approach was used recently to several
high-energy reactions, including heavy quark pair photo-
\cite{LS04,Mariotto} and hadroproduction \cite{BS00,LS06},
charmonium production \cite{HKSST1,HKSST2}, inclusive $Z_0$
\cite{KS04} and Higgs \cite{LZ05,LS05} production.

In the present paper we shall compare results obtained in the
leading-order k$_t$-factorization approach and the next-to-leading order
collinear-factorization approach.
We shall discuss which approach is more adequate for the different
regions of phase space. We shall present corresponding results for
proton-proton scattering at RHIC and proton-antiproton scattering
at Tevatron.

\begin{widetext}
%===========================
\section{Formalism}
%===========================

%--------------------------------------------------------------------------
\subsection{$2 \to 2$ contributions with unintegrated parton
distributions}
%--------------------------------------------------------------------------

It is known that at midrapidities and at relatively small transverse momenta
the photon-jet production is dominated by (sub)processes initiated by gluons.
In Fig.\ref{fig:2to2_diagrams} we show basic diagrams which appear 
in the $k_t$-factorization approach to photon-jet correlations.
%------------------------------
\begin{figure}[!htb] % Figure 1
\begin{center}
\subfigure[]{\label{nlo_a}
\includegraphics[width=4cm]{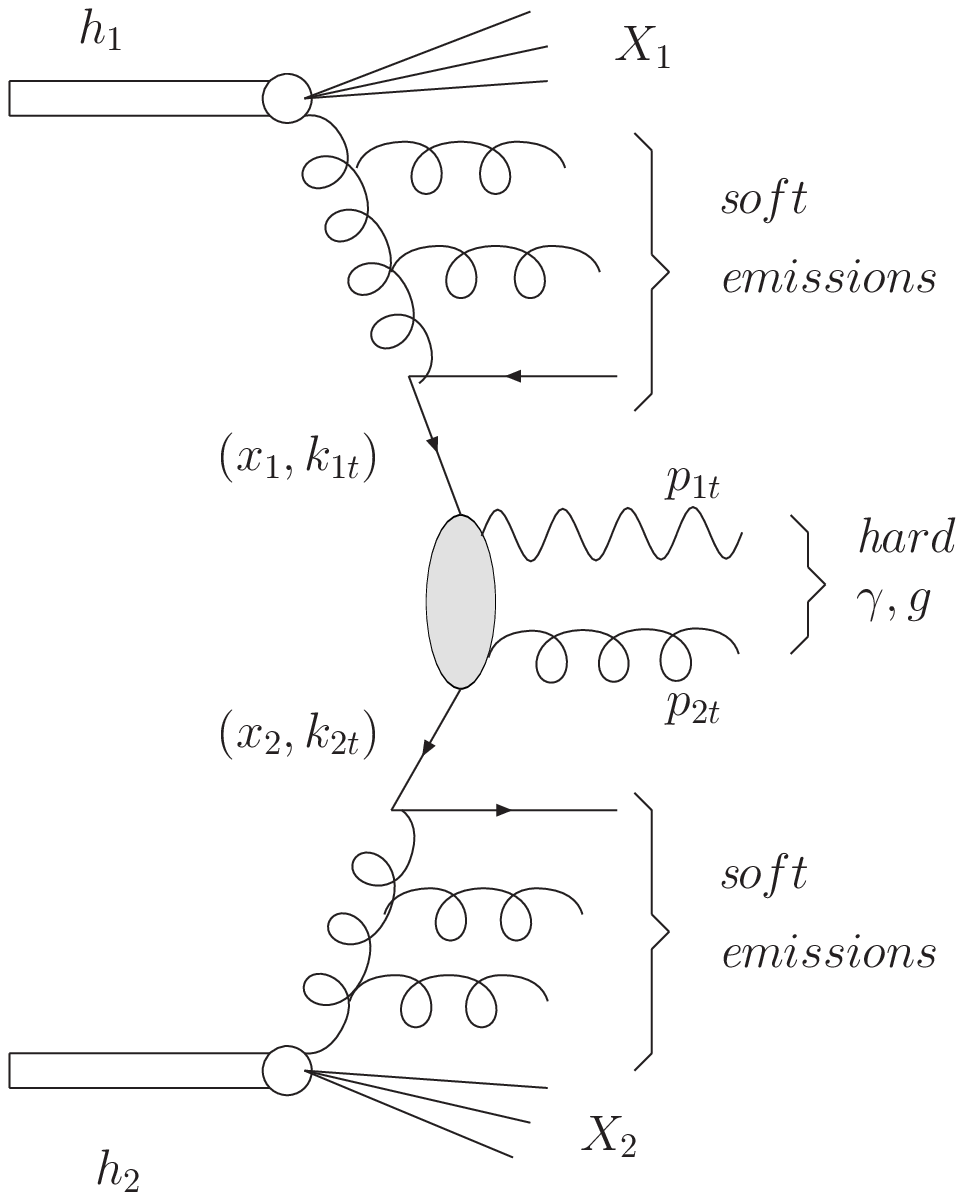}}
\subfigure[]{\label{nlo_a}
\includegraphics[width=4cm]{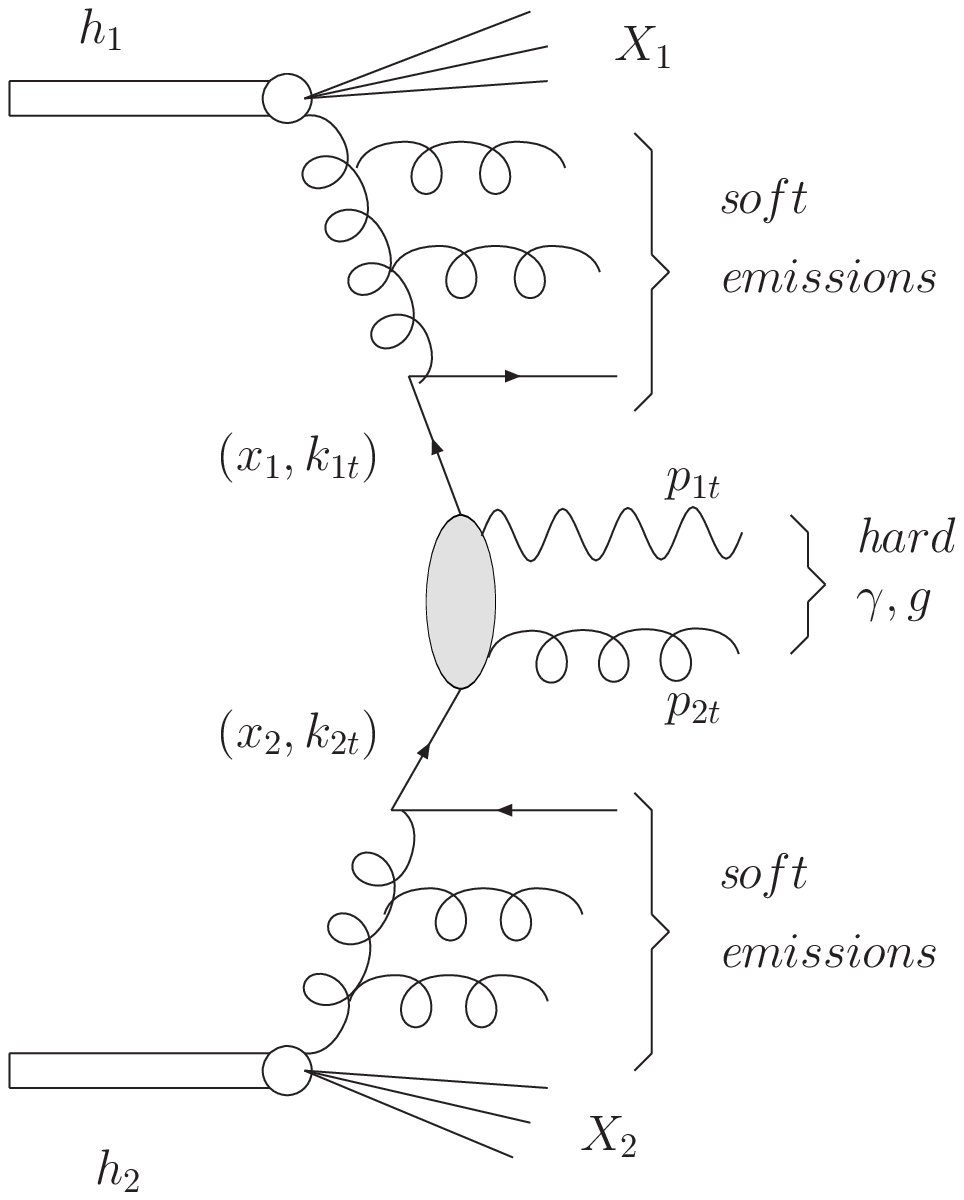}}
\subfigure[]{\label{nlo_a}
\includegraphics[width=4cm]{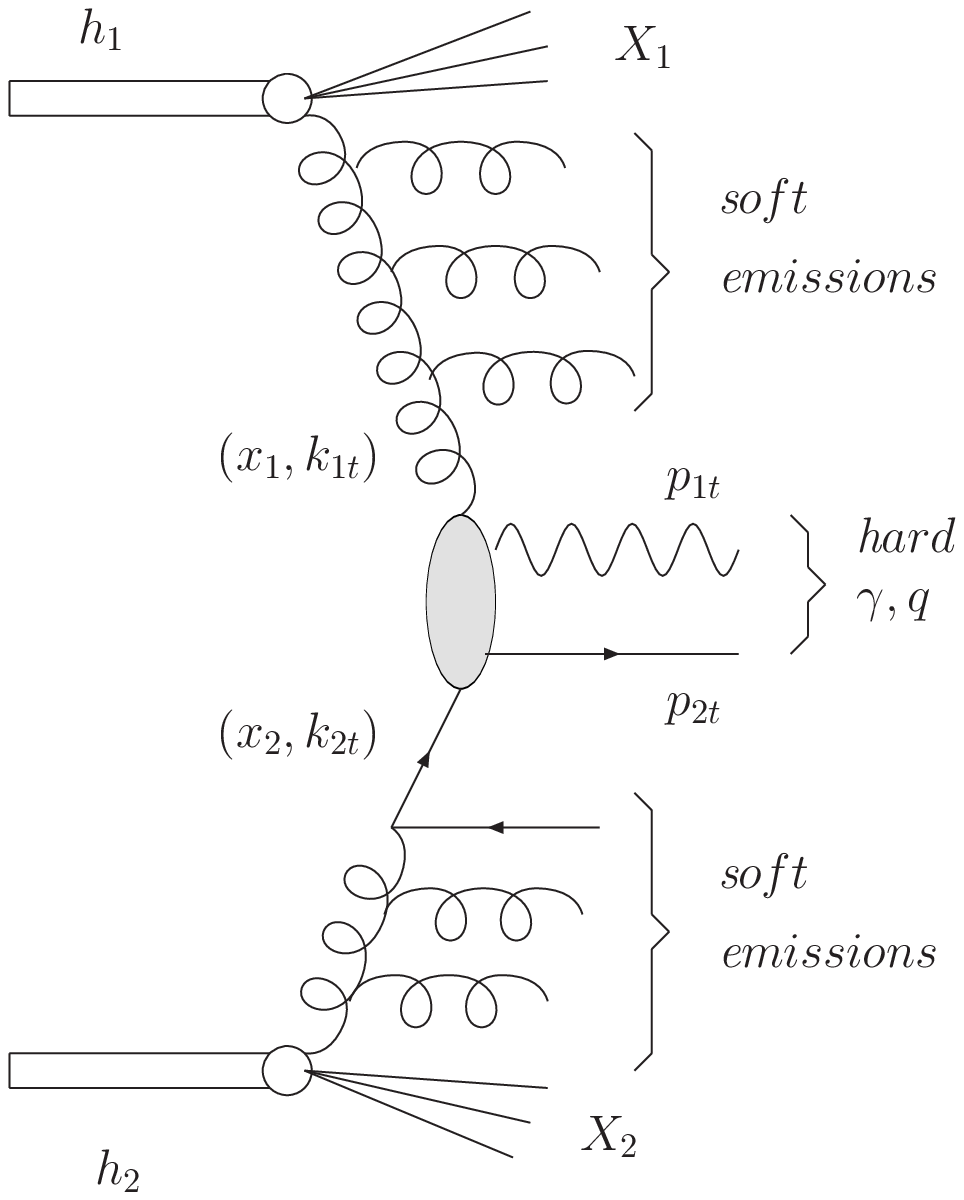}}
\subfigure[]{\label{nlo_a}
\includegraphics[width=4cm]{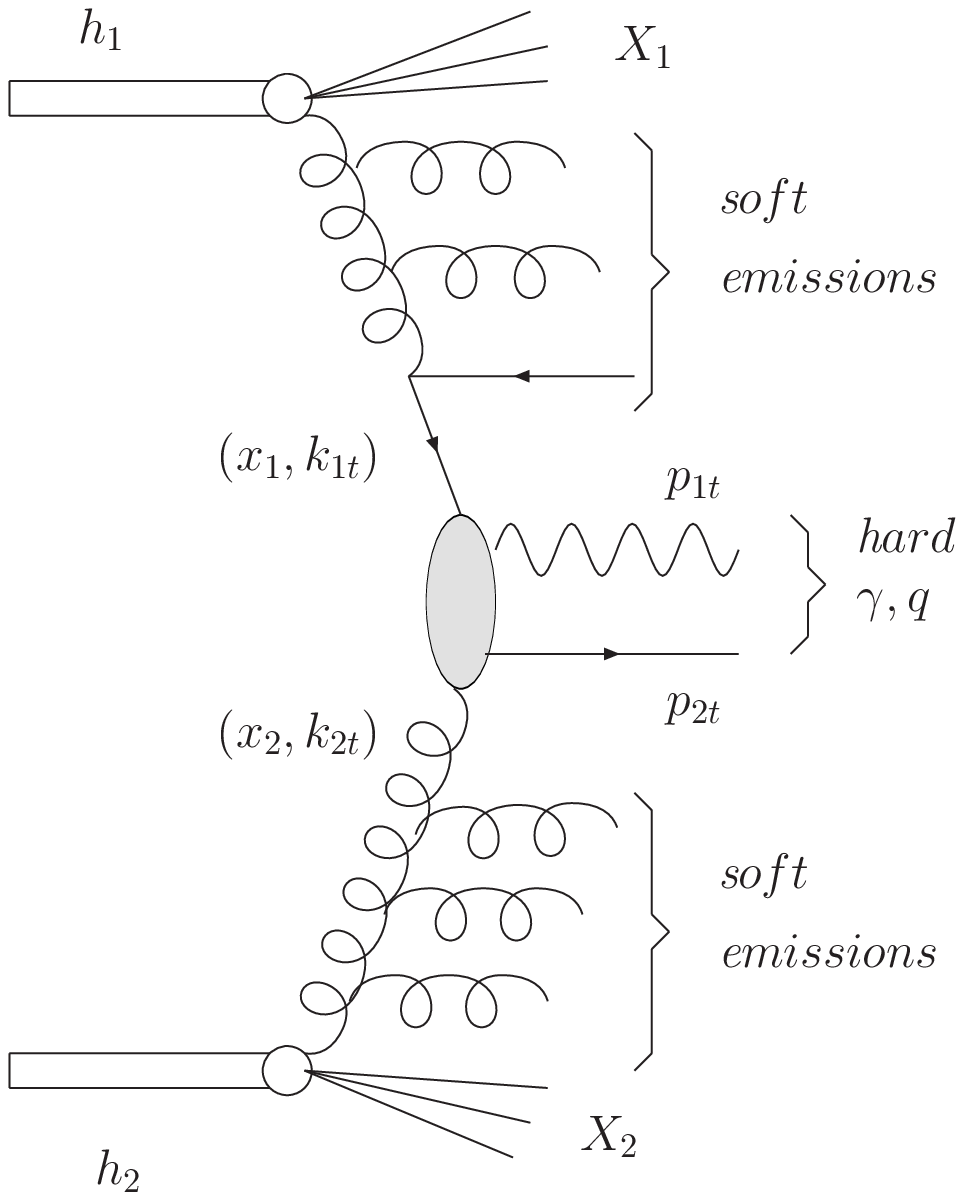}}
\caption{
Basic diagrams of $k_t$-factorization approach to photon-jet correlations.
\label{fig:2to2_diagrams}
}
\end{center}
\end{figure}
%-----------
%In this paper we concentrate only on such processes.

In the $k_t$-factorization approach the cross section for a simultaneous
production of a photon and an associated jet in the collisions of two hadrons
($pp$ or $p \bar p$) can be written as
\begin{eqnarray} 
\frac{d\sigma_{h_1h_2 \to \gamma k}}{d^2p_{1,t}d^2p_{2,t}}
&=& \int dy_1dy_2 \frac{d^2k_{1,t}}{\pi}\frac{d^2k_{1,t}}{\pi}
\frac{1}{16\pi^2(x_1x_2s)^2}\overline{|{\mathcal M}_{ij\to \gamma k}|^2}\nonumber \\
&\cdot& \delta^2(\vec{k}_{1,t}+\vec{k}_{2,t}-\vec{p}_{1,t}-\vec{p}_{2,t})
{\cal F}_i(x_1,k_{1,t}^2,\mu_1^2) {\cal F}_j(x_2,k_{2,t}^2,\mu_2^2)  \; ,
\label{photon_jet_cross_section}
\end{eqnarray} 
where ${\cal F}_i(x_1,k_{1,t}^2,\mu_1^2)$ and ${\cal F}_j(x_2,k_{2,t}^2,\mu_2^2)$
are so-called unintegrated parton distributions.
The longitudinal momentum fractions are evaluated as
\begin{eqnarray}
x_1=(m_{1t}{\mathrm e}^{+y_1}&+&m_{2t}{\mathrm e}^{+y_2})/\sqrt{s}\; , \nonumber \\
x_2=(m_{1t}{\mathrm e}^{-y_1}&+&m_{2t}{\mathrm e}^{-y_2})/\sqrt{s}\; . 
\label{x1_x2}
\end{eqnarray}
We shall return to the choice of the factorization scale in the next
section. Its role is completely different in different approaches
i.e. different choices of UPDFs. A special attention will be devoted to
the Kwieci\'nski UPDF and the role of the scale paramater. 
 
If one makes the following replacement 
\begin{eqnarray}
{\cal F}_i(x_1,k_{1,t}^2) &\rightarrow& x_1p_i(x_1)\delta (k_{1,t}^2)
\nonumber \;, \\
{\cal F}_j(x_2,k_{2,t}^2) &\rightarrow& x_2p_j(x_2)\delta (k_{2,t}^2)
\label{f1_f2}
\end{eqnarray}
then one recovers the standard leading-order collinear formula.

The final partonic state is $\gamma k = \gamma g, \gamma q$. 
The matrix elements for corresponding processes are discussed in
Appendix A.

The inclusive invariant cross section for direct photon production can
be written as
\begin{eqnarray}
\frac{d\sigma_{h_1h_2\to \gamma}}{dy_1d^2p_{1,t}} &=& \int dy_2
\frac{d^2k_{1,t}}{\pi}\frac{d^2k_{2,t}}{\pi}(...)|_{\vec{p}_{2,t}=\vec{k}_{1,t}+
\vec{k}_{2,t}-\vec{p}_{1,t}}
\label{inclusive_photon}
\end{eqnarray}
and analogously the cross section for the associated parton (jet) can
be written as
\begin{eqnarray}
\frac{d\sigma_{h_1h_2\to k}}{dy_1d^2p_{1,t}} &=& \int dy_1
\frac{d^2k_{1,t}}{\pi}\frac{d^2k_{2,t}}{\pi}(...)|_{\vec{p}_{1,t}=\vec{k}_{1,t}+
\vec{k}_{2,t}-\vec{p}_{2,t}} \; .
\label{inclusive_jet}
\end{eqnarray}
Let us return to the coincidence cross section. The integration 
with the Dirac delta function in Eq.(\ref{photon_jet_cross_section})
\begin{eqnarray}
\int dy_1dy_2\frac{d^2k_{1,t}}{\pi}\frac{d^2k_{2,t}}{\pi}
(...)\delta^2(...)
\label{coincidence_cross_section}
\end{eqnarray}
can be performed by introducing the following new auxiliary variables:
\begin{eqnarray}
\vec{Q}_t &=& \vec{k}_{1,t}+\vec{k}_{2,t}\; , \nonumber \\
\vec{q}_t &=& \vec{k}_{1,t}-\vec{k}_{2,t}\; .
\label{q_t}
\end{eqnarray}
Then our initial cross section can be written as:
\begin{eqnarray}
\frac{d\sigma_{h_1h_2 \to \gamma , parton}}{d^2p_{1,t}d^2p_{2,t}}
=\frac{1}{4}\int dy_1dy_2 \; \frac{1}{2}dq_t^2 d\phi_{q_t} (...)
|_{\vec{Q}_t = \vec{P}_t}\; .
\label{photon_jet_in_qt}
\end{eqnarray}
Above $\vec{P}_t = \vec{p}_{1,t}+\vec{p}_{2,t}$. The factor $\frac{1}{4}$
before integrand on the rhs comes from the Jacobian of the 
$(\vec{k}_{1,t}, \vec{k}_{2,t}) \to (\vec{Q}_t, \vec{q}_t)$ transformation
(see \cite{PS06_photon}).

%-----------------------------------------------------------------------
\subsection{$2 \to 3$ contributions in NLO collinear-factorization approach}
%-----------------------------------------------------------------------

Up to now we have concentrated only on processes with two explicit 
hard partons ($\gamma k$) in the $k_t$-factorization approach.
It is of interest to compare the results of our approach with those of
the standard collinear next-to-leading order approach.
In this section we discuss processes with three explicit hard partons.
In Fig.\ref{fig:nlo_diagrams} we show diagrams for $2 \to 3$
subprocesses included in our calculations.
In the following we assume particle No.1 to be a photon. Then
particle No.2 is $g, q$ and $\bar q$, depending on the subprocess.

%-----------------------------
\begin{figure}[htb] % Figure 2
\begin{center}
\subfigure[]{\label{nlo_a}
\includegraphics[height=2cm]{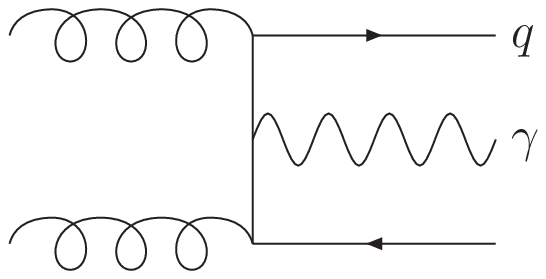}}
\subfigure[]{\label{nlo_b}
\includegraphics[height=2cm]{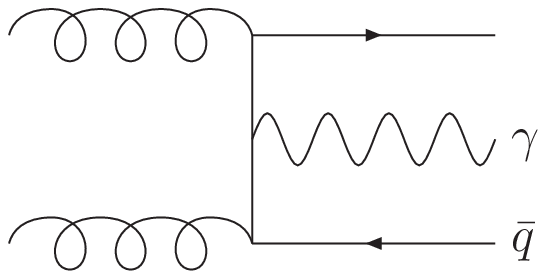}}
\subfigure[]{\label{nlo_c}
\includegraphics[height=2cm]{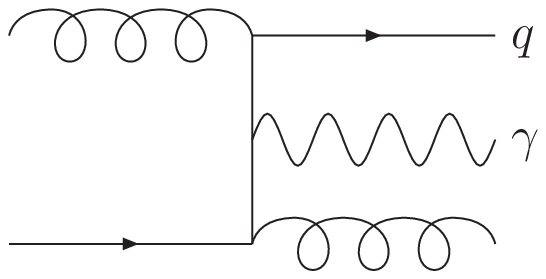}}
\subfigure[]{\label{nlo_d}
\includegraphics[height=2cm]{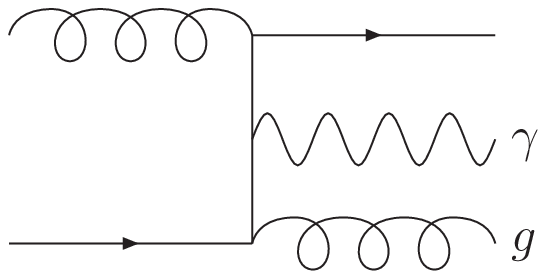}}
\subfigure[]{\label{nlo_e}
\includegraphics[height=2cm]{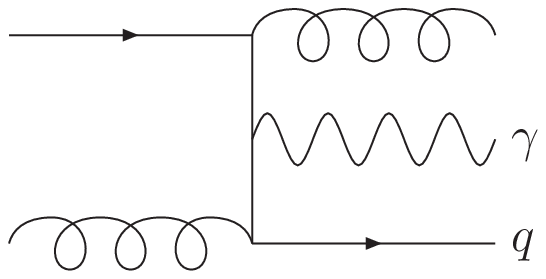}}
\subfigure[]{\label{nlo_f}
\includegraphics[height=2cm]{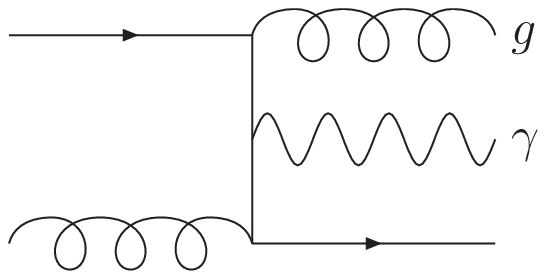}}
\subfigure[]{\label{nlo_g}
\includegraphics[height=2cm]{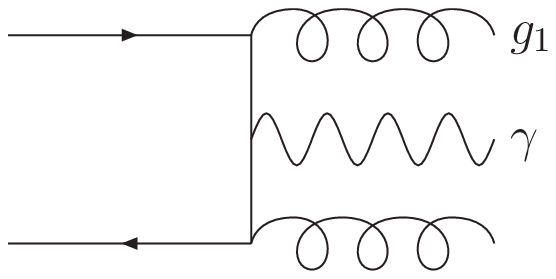}}
\subfigure[]{\label{nlo_h}
\includegraphics[height=2cm]{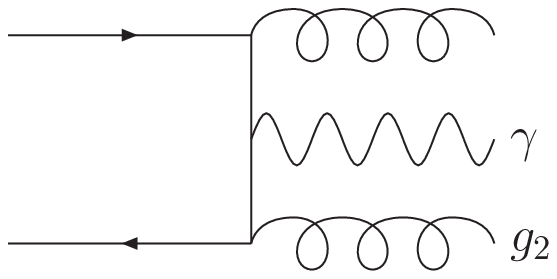}}
\caption{
Diagrams for NLO collinear-factorization approach for photon-jet-jet
production.
\label{fig:nlo_diagrams}
}
\end{center}
\end{figure}
%-----------

The cross section for $h_1 h_2\to \gamma k l X$  processes can be calculated
according to the standard parton model formula
\begin{eqnarray}
d\sigma_{h_1h_2\to \gamma kl} &=& \sum_{ijk}
\int dx_1dx_2 \; p_i(x_1,\mu^2)p_j(x_2,\mu^2) \;
d\hat{\sigma}_{ij\to \gamma kl}    \; . 
\label{dsigma_h1h2}
\end{eqnarray}
The elementary cross section can be written as
\begin{eqnarray}
d\hat{\sigma}_{ij\to \gamma kl} &=& \frac{1}{2\hat s}
\overline{|{\mathcal M}_{ij\to \gamma kl}|^2} dR_3 \; ,
\label{dsigma_ij}
\end{eqnarray}
where the three-body phase space element reads
\begin{eqnarray}
dR_3&=&(2\pi)^4\delta^4(p_a + p_b - \sum_{i=1}^3 p_i) 
\prod_{i=1}^3\frac{d^3p_i}{2E_i(2\pi)^3} \; .
\label{dR_3a}
\end{eqnarray}
This element can be expressed in an equivalent way in terms of parton 
rapidities
\begin{eqnarray}
dR_3&=&(2\pi)^4\delta^4(p_a + p_b - \sum_{i=1}^3 p_i) 
\prod_{i=1}^3\frac{dy_id^2p_{i,t}}{(4\pi)(2\pi)^2} \; .
\label{dR_3b}
\end{eqnarray}
The last formula is useful for practical applications. Now the cross section
for hadronic collisions can be written in terms of $2 \to 3$ matrix
element as 
\begin{eqnarray}
d\sigma &=& \sum_{ijkl}
dy_1d^2p_{1,t}dy_2d^2p_{2,t}dy_3 \frac{1}{(4\pi)^3(2\pi)^2}\frac{1}{\hat s^2}
x_1p_i(x_1,\mu^2)x_2p_j(x_2,\mu^2)\overline{|{\mathcal M}_{ij \to \gamma
  k l}|^2} \; ,
\label{dsigma}
\end{eqnarray}
where the longitudinal momentum fractions are evaluated as
\begin{eqnarray}
x_1&=& \frac{1}{\sqrt s} \sum_{i=1}^3 p_{i,t}{\mathrm e}^{+y_i} \; ,
\nonumber \\
x_2&=& \frac{1}{\sqrt s} \sum_{i=1}^3 p_{i,t}{\mathrm e}^{-y_i} \; .
\label{x1x2_2to3}
\end{eqnarray}
Repeating similar steps as for $2 \to 2$ processes we get finally
\begin{eqnarray}
d\sigma&=&
\sum_{ijkl}
\frac{1}{64\pi^4\hat s^2}x_1p_i(x_1,\mu^2)x_2p_j(x_2,\mu^2)
\overline{|{\mathcal M}_{ij \to \gamma k l}|^2}\;p_{1,t}dp_{1,t}p_{2,t}dp_{2,t}
d\phi_-dy_1dy_2dy_3 \; , 
\label{2to3_useful}
\end{eqnarray}
where the relative azimuthal angle between the photon and the associated jet
($\phi_-$) is restricted to the interval $(0,\pi)$. The last formula 
is very useful in calculating the cross section for 
particle 1 and particle 2 correlations.

%======================
\section{Results}
%======================

In this section we shall present results for RHIC and Tevatron energies.
We use UPDFs from the literature.
There are only two complete sets of UPDF in the literature
which include not only the gluon distributions but also the distributions
of quarks and antiquarks:\\
(a) Kwieci\'nski \cite{Kwiecinski},\\
(b) Kimber-Martin-Ryskin \cite{KMR}.\\
For comparison we shall include also the unintegrated parton distributions
obtained from the collinear ones by the Gaussian smearing procedure.
Such a procedure is often used in the context of direct photons
\cite{Owens,AM04}.
Comparing results obtained with those Gaussian distributions
and the results obtained with the Kwieci\'nski distributions with
nonperturbative Gaussian form factors will allow
to quantify the effect of UPDF evolution as contained in the
Kwieci\'nski evolution equations. What is the hard scale for our process?
In our case the best candidate for the scale
is the photon and/or jet transverse momentum. Since we are interested
in rather small transverse momenta the evolution length is not
too large and the deviations from initial $k_t$-distributions
(assumed here to be Gaussian) should not be too big.

At high energies one enters into a small-x region, i.e. the region
of a specific dynamics of the QCD emissions. In this region only unintegrated
distributions of gluons exist in the literature. In our case
the dominant contributions come from QCD-Compton
$gluon-quark$ or $quark-gluon$ initiated hard subprocesses. 
This means that we need unintegrated distributions of both
gluons and quarks/antiquarks. In this case we take such UGDFs from the
literature and supplement them by the Gaussian distributions of 
quarks/antiquarks.

Let us start from presenting our results on the $(p_{1,t},p_{2,t})$ plane.
In Fig.\ref{fig:updfs_ptpt} we show the maps for different 
UPDFs used in the $k_t$-factorization approach as well as for NLO 
collinear-factorization approach for
$p_{1,t}, p_{2,t} \in (5,20)$~GeV and at the Tevatron energy $W =
1960$~GeV. In the case of the Kwieci\'nski distribution we have taken
$b_0$ = 1 GeV$^{-1}$ for the exponential nonperturbative form factor
and the scale parameter $\mu^2$ = 100 GeV$^2$.
Rather similar distributions are obtained for different UPDFs.
The distribution obtained in the NLO approach differs qualitatively
from those obtained in the $k_t$-factorization approach.
First of all, one can see a sharp ridge along the diagonal $p_{1,t} = p_{2,t}$.
This ridge corresponds to a soft singularity when the unobserved
parton has very small transverse momentum $p_{3,t}$.
As will be clear in a moment this corresponds to the azimuthal
angle between the photon and the jet being $\phi_{-} = \pi$. Obviously this is
a region which cannot be reliably calculated in collinear pQCD.
There are different practical possibilities to exclude this region from
the calculations.
The most primitive way (possible only in theoretical calculations) is to
impose a lower cut on transverse momentum of the unobserved parton $p_{3,t}$.
Secondly, the standard collinear NLO approach generates much bigger
cross section at configurations asymmetric in $p_{1,t}$ and $p_{2,t}$.
We shall return to this observation in the course of this paper. 

%------------------------------------------------------------------
\begin{figure}[!htb] % Figure 3
\begin{center}
\subfigure[]{\label{a}
\includegraphics[height=6cm]{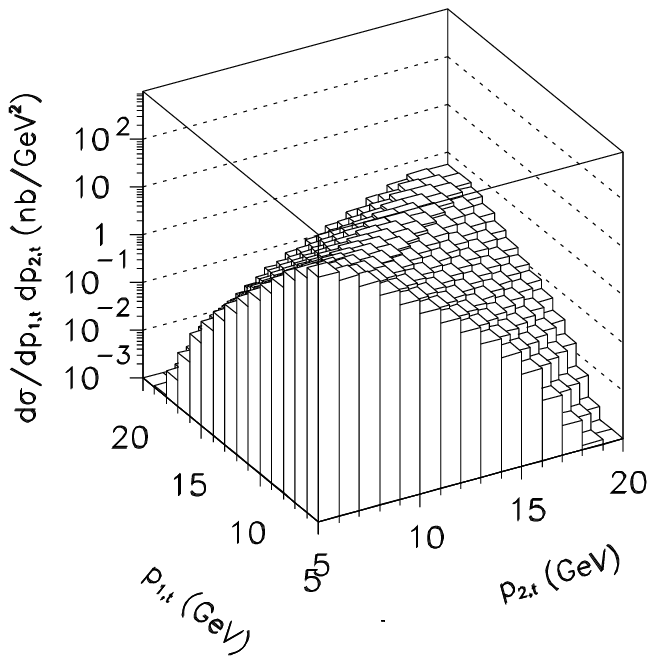}}
\subfigure[]{\label{a}
\includegraphics[height=6cm]{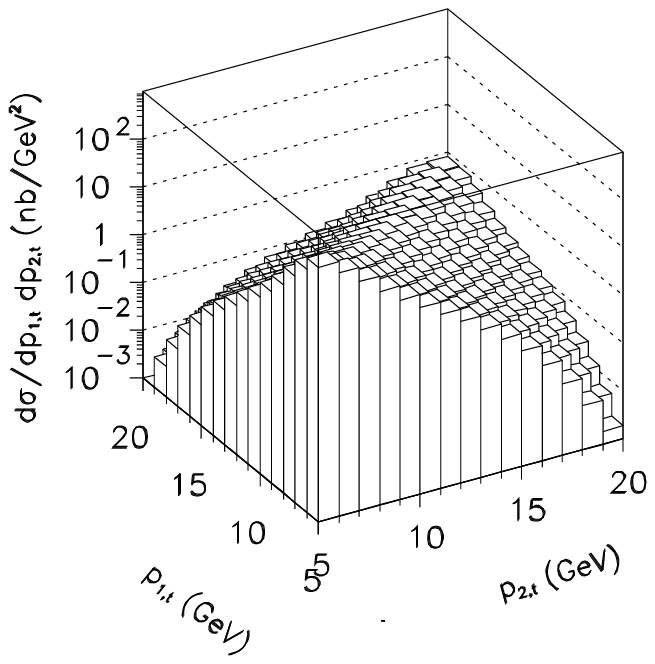}}
\subfigure[]{\label{a}
\includegraphics[height=6cm]{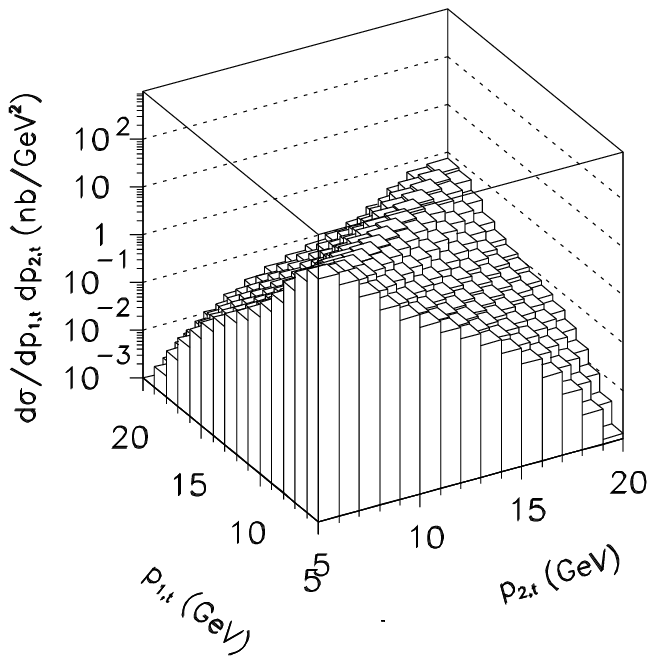}}
\subfigure[]{\label{a}
\includegraphics[height=6cm]{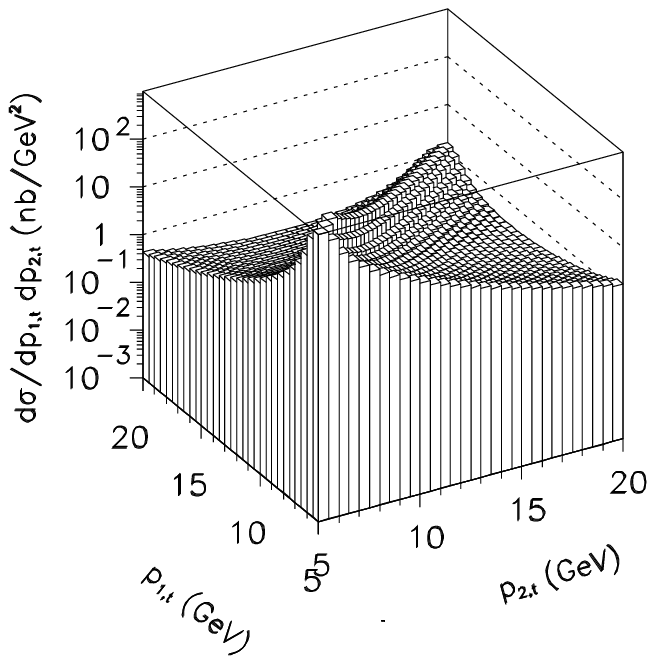}}
\caption{
Transverse momentum distributions $d\sigma/dp_{1,t}dp_{2,t}$
at W = 1960 GeV and  
for different UPDFs in the $k_t$-factorization approach for
Kwieci\'nski ($b_0 = 1 GeV^{-1}$, $\mu^2 = 100 GeV^{2}$) (a), BFKL (b), KL
(c) and NLO $2 \to 3$ collinear-factorization approach including
diagrams from Fig.\ref{fig:nlo_diagrams} (d).
The integration over rapidities from the interval -5 $< y_1, y_2 <$
5 is performed.
\label{fig:updfs_ptpt}
}
\end{center}
\end{figure}
%------------------------------------------------------------------

As discussed in Ref.\cite{PS06_photon} the Kwieci\'nski distributions
are very useful to treat both the nonperturbative (intrinsic
nonperturbative transverse momenta)
and the perturbative (QCD broadening due to parton emission) effects on
the same footing.
In Fig.\ref{fig:kwiecinski_scale} we show the effect of
the scale evolution of the Kwieci\'nski UPDFs on the azimuthal angle
correlations between the photon and the associated jet.
We show results for different initial conditions ($b_0$ = 0.5, 1.0, 2.0
GeV$^{-1}$). At the initial scale (fixed here as in the original
GRV \cite{GRV98} to be $\mu^2$ = 0.25 GeV$^2$) there is a sizeable
difference of the results for different $b_0$. The difference
becomes less and less pronounced when the scale increases.
At $\mu^2$ = 100 GeV$^2$ the differences practically disappear.
This is due to the fact that the QCD-evolution broadening of
the initial parton transverse momentum distribution is much bigger than
the typical initial nonperturbative transverse momentum scale.

%------------------------------
\begin{figure}[!htb] % Figure 4
\begin{center}
\subfigure[]{\label{a}
\includegraphics[height=6cm]{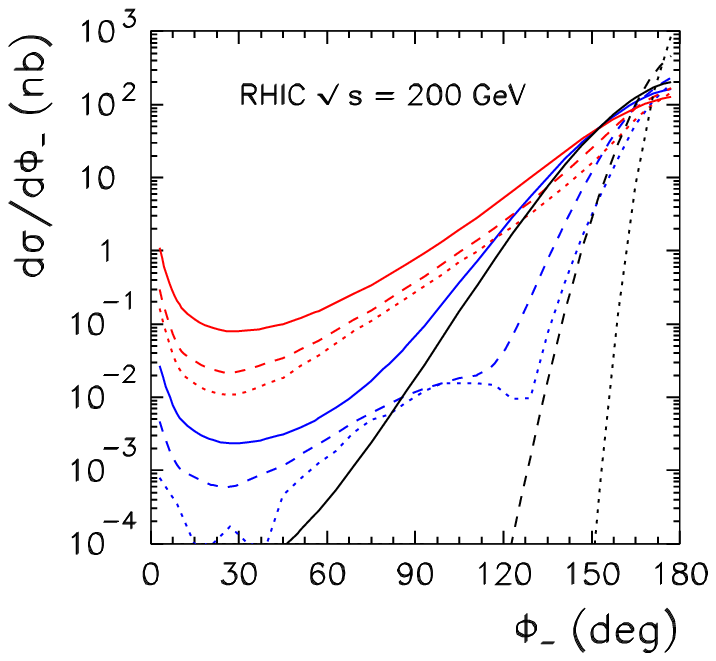}}
\subfigure[]{\label{a}
\includegraphics[height=6cm]{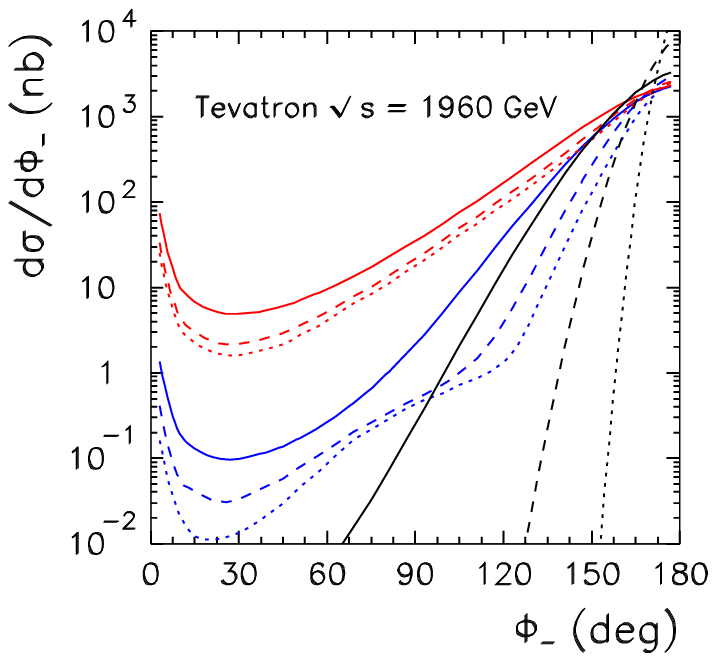}}
\caption{
(Color on line) Azimuthal angle correlation functions at (a) RHIC, (b) Tevatron
energies for different scales and different values of $b_0$ 
of the Kwieci\'nski distributions.
The solid line is for $b_0$ = 0.5 GeV$^{-1}$, the dashed line is for
$b_0$ = 1 GeV$^{-1}$ and the dotted line is for $b_0$ = 2 GeV$^{-1}$.
Three different values of the scale parameters are shown: 
$\mu^2$ = 0.25, 10, 100 GeV$^2$ (the bigger the scale the bigger
the decorellation effect, different colors on line).
In this calculation  $p_{1,t}, p_{2,t} \in$ (5,20) GeV and
$y_1, y_2 \in$ (-5,5).
\label{fig:kwiecinski_scale}
}
\end{center}
\end{figure}
%-----------

In Fig.\ref{fig:updfs_phid} we show corresponding azimuthal angular
correlations.
% again for different
%UPDFs in the $k_t$-factorization approach and NLO collinear-factorization
%approach.for the Tevatron energy $W = 1960$~GeV.
In this case integration is made over transverse momenta 
$p_{1,t}, p_{2,t} \in (5,20)$~GeV and rapidities $y_1, y_2 \in
(-5,5)$.

%---------------------------------------------------------------------
\begin{figure}[!htb] % Figure 5 
\begin{center}
\includegraphics[height=6cm]{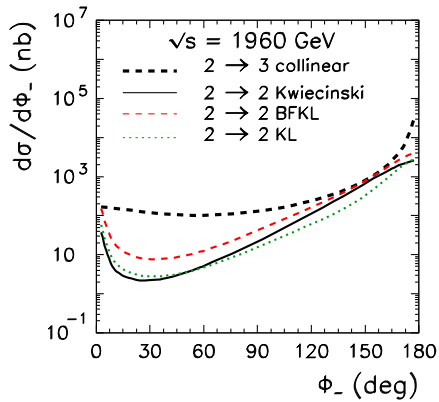}
\caption{
Photon-jet angular azimuthal correlations $d\sigma/d\phi_-$
for proton-antiproton collision at W = 1960 GeV  
for different UPDFs in the $k_t$-factorization approach for
the Kwieci\'nski (solid), BFKL (dashed), KL (dotted) UPDFs/UGDFs
and for the NLO collinear-factorization approach (thick dashed).
Here $y_1, y_2 \in (-5,5)$.
\label{fig:updfs_phid}
}
\end{center}
\end{figure}
%---------------------------------------------------------------------

The singularity in NLO pQCD at $\phi_{-} = \pi$ is strongly correlated
with the sharp ridge in Fig.\ref{fig:updfs_ptpt} d. This is demonstrated in 
Fig.\ref{fig:NLO_p3t_cut} where we present the results of azimuthal
correlation function obtained for different cuts on $p_{3,t}$. The cut
modifies only the region of relative azimuthal angles close to $\pi$.
We wish to stress in this context that there are no singularities
of the ridge type in the $k_t$-factorization approach.

%---------------------------------------------------------------------
\begin{figure}[!htb] % Figure 6
\begin{center}
\includegraphics[height=6cm]{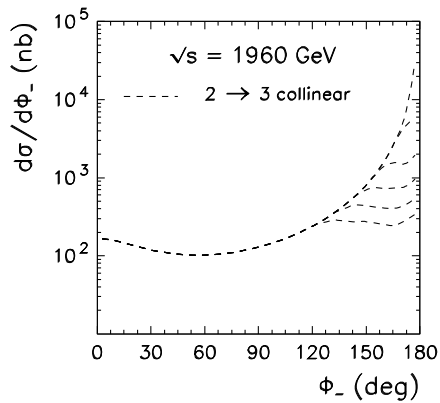}
\caption{
Photon-jet angular azimuthal correlations $d\sigma/d\phi_-$
for proton-antiproton collision at W = 1960 GeV  
for the NLO collinear-factorization approach and different cuts on
$p_{3,t}$. Here $y_1, y_2 \in (-5,5)$.
\label{fig:NLO_p3t_cut}
}
\end{center}
\end{figure}
%---------------------------------------------------------------------
 
The small transverse momenta of the unobserved jet contribute to the sharp
ridge along the diagonal $p_{1,t} = p_{2,t}$. It is therefore difficult
to distinguish these three-parton states from the states with two partons.
The ridge can be eliminated in calculation by imposing a cut on the
transverse momentum of the third (unobserved) parton. 
In experiments there is no possibility to impose such
cuts and other methods must be used. We shall return to this point later
in this paper.%
%\begin{eqnarray*}
%\Theta_2 (p_{2,t} &-& p_{3,t}) \\
%\Theta_1 (p_{1,t} &-& p_{3,t})
%\label{}
%\end{eqnarray*}
%

In Fig.\ref{fig:leading_jets_angle} we show angular azimuthal
correlations for different relations between transverse momenta
of outgoing photon and partons:
(a) with no constraints on $p_{3,t}$, (b) the case where $p_{2,t} >
p_{3,t}$ condition (called leading jet condition in the following)
is imposed, (c) $p_{2,t} > p_{3,t}$ and
an additional condition $p_{1,t} > p_{3,t}$.
The results depend significantly on the scenario as can be seen
from the figure. 

%------------------------------
\begin{figure}[!htb] % Figure 7
\begin{center}
\includegraphics[height=6cm]{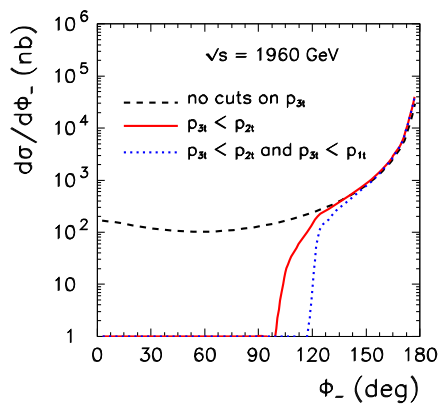}
\caption{
Angular azimuthal correlations for different cuts on the transverse 
momentum of third (unobserved) parton in the NLO collinear-factorization approach
without any extra constraints (dashed), $p_{3,t} < p_{2,t}$ (solid),
$p_{3,t} < p_{2,t}$ and $p_{3,t} < p_{1,t}$ in addition (dotted).
Here $y_1, y_2 \in (-5,5)$.
\label{fig:leading_jets_angle}
}
\end{center}
\end{figure}
%------------------------------

In Fig.\ref{fig:leading_jets_ptpt} we show transverse momentum distribution 
$d\sigma/dp_{1,t}dp_{2,t}$ for the same extra conditions imposed
before in Fig.\ref{fig:leading_jets_angle} for azimuthal angle correlations.
Imposing the condition that the associated jet is the leading jet
$(p_{2,t} > p_{3,t})$ causes that the large part of the phase space
$p_{1,t} < 2 p_{2,t}$ is not available in the next-to-leading approach.
If one imposes in addition that $p_{1,t}$(photon) $>$ $p_{3,t}$(unobserved jet)
then also the $p_{2,t} < 2 p_{1,t}$ region becomes excluded for the NLO
approach. These NLO-excluded regions are therefore regions sensitive
to higher-order corrections in pQCD.

%------------------------------
\begin{figure}[!htb] % Figure 8
\begin{center}
\subfigure[]{\label{a}
\includegraphics[height=6cm]{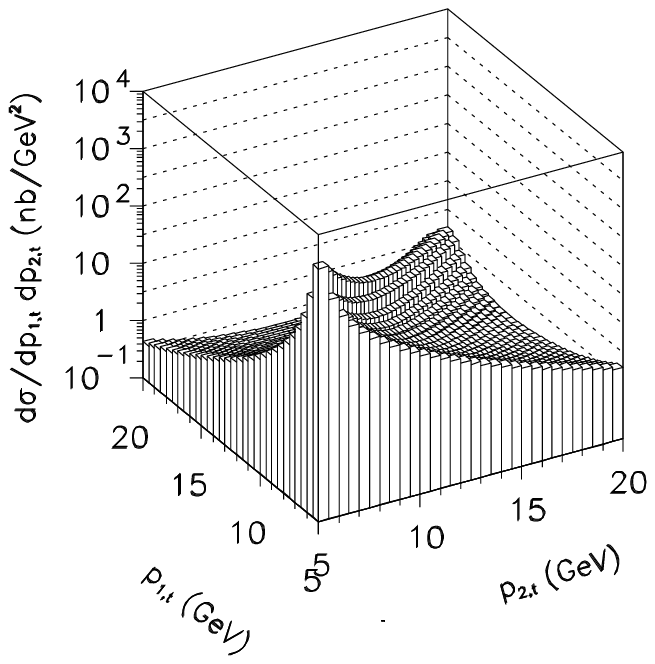}}
\subfigure[]{\label{a}
\includegraphics[height=6cm]{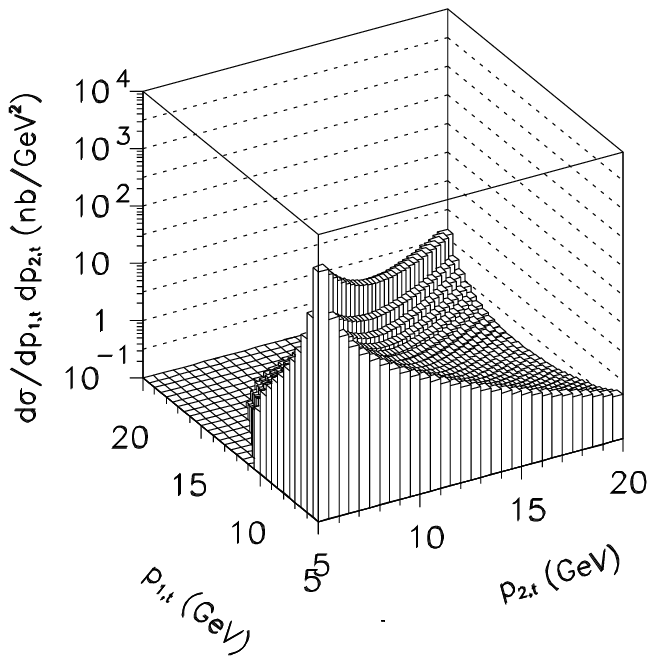}}
\subfigure[]{\label{a}
\includegraphics[height=6cm]{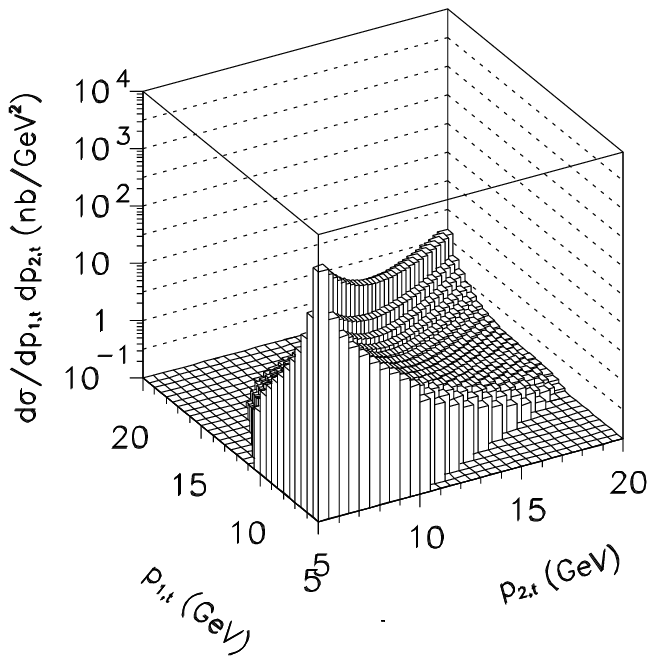}}
\caption{
Transverse momenta distribution $d\sigma/dp_{1,t}dp_{2,t}$ for 
different constraints on the transverse momentum 
of third (unobserved) parton in the NLO collinear-factorization approach
with no constraints on $p_{3,t}$ (a), $p_{3,t} < p_{2,t}$ (b),
$p_{3,t} < p_{2,t}$ and $p_{3,t} < p_{1,t}$ (c).
All $2 \to 3$ processes shown in Fig.\ref{fig:nlo_diagrams} were
included.
Here $y_1, y_2 \in (-5,5)$.
\label{fig:leading_jets_ptpt}
}
\end{center}
\end{figure}
%-----------

In general, the correlations between the photon and the jet depend strongly
on all kinematical variables - transverse momenta, azimuthal angles, etc.
In order to expose this better, in Fig.\ref{fig:ptpt_planes} we define
windows in the $(p_{1,t}, p_{2,t})$ plane which will be used in the
following to study the azimuthal correlations.
At lower energies (as for RHIC) a region of rather low transverse
momenta is more adequate (left figure). At larger energies (as for
Tevatron) also region of somewhat larger transverse momenta can be of
interest (right figure). The notation shown in the figure will be
used for brevity in the rest of this paper for easy reference.

%------------------------------
\begin{figure}[!htb] % Figure 9
\begin{center}
\subfigure[]{\label{a}
\includegraphics[height=6cm]{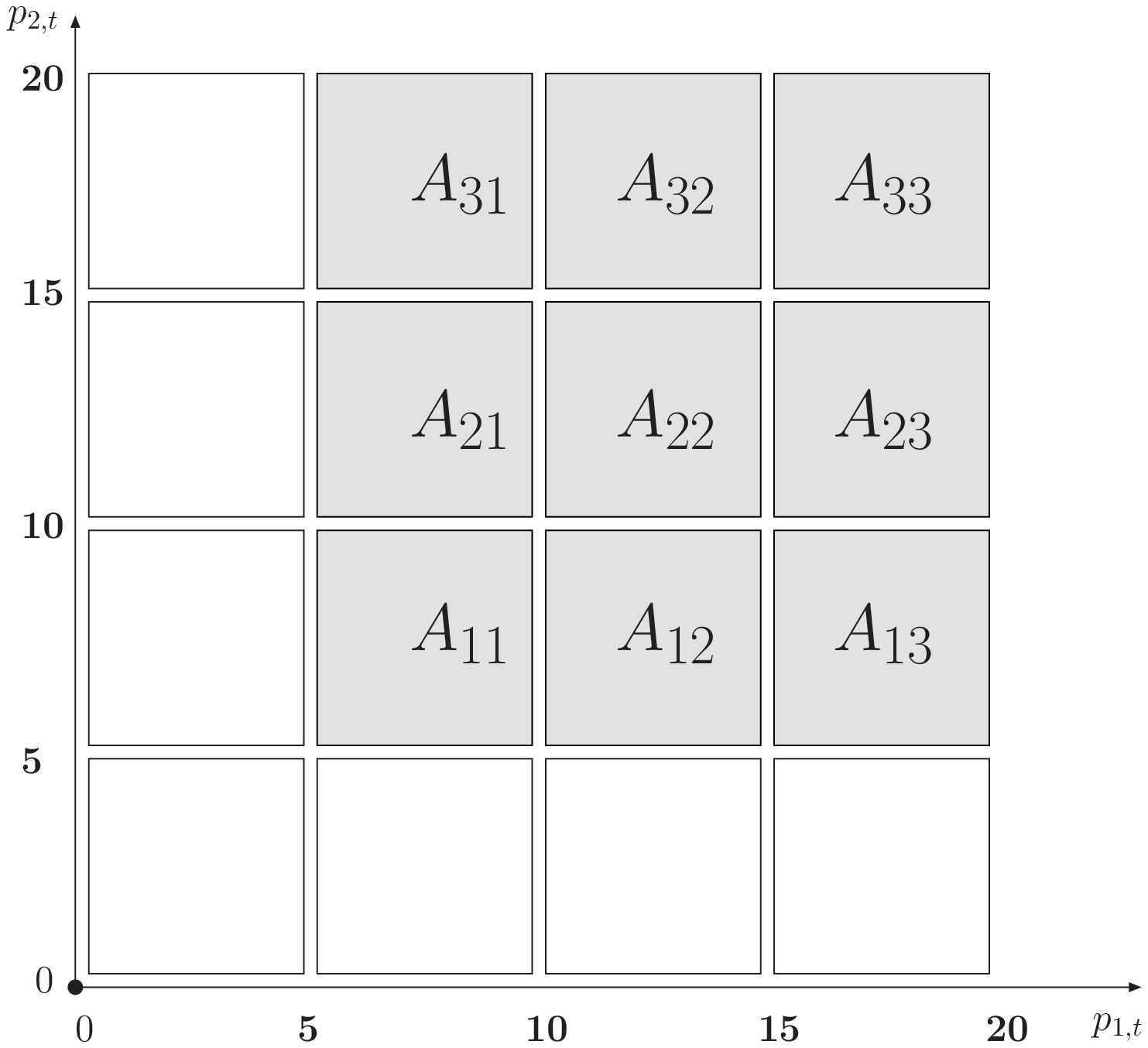}}
\subfigure[]{\label{a}
\includegraphics[height=6cm]{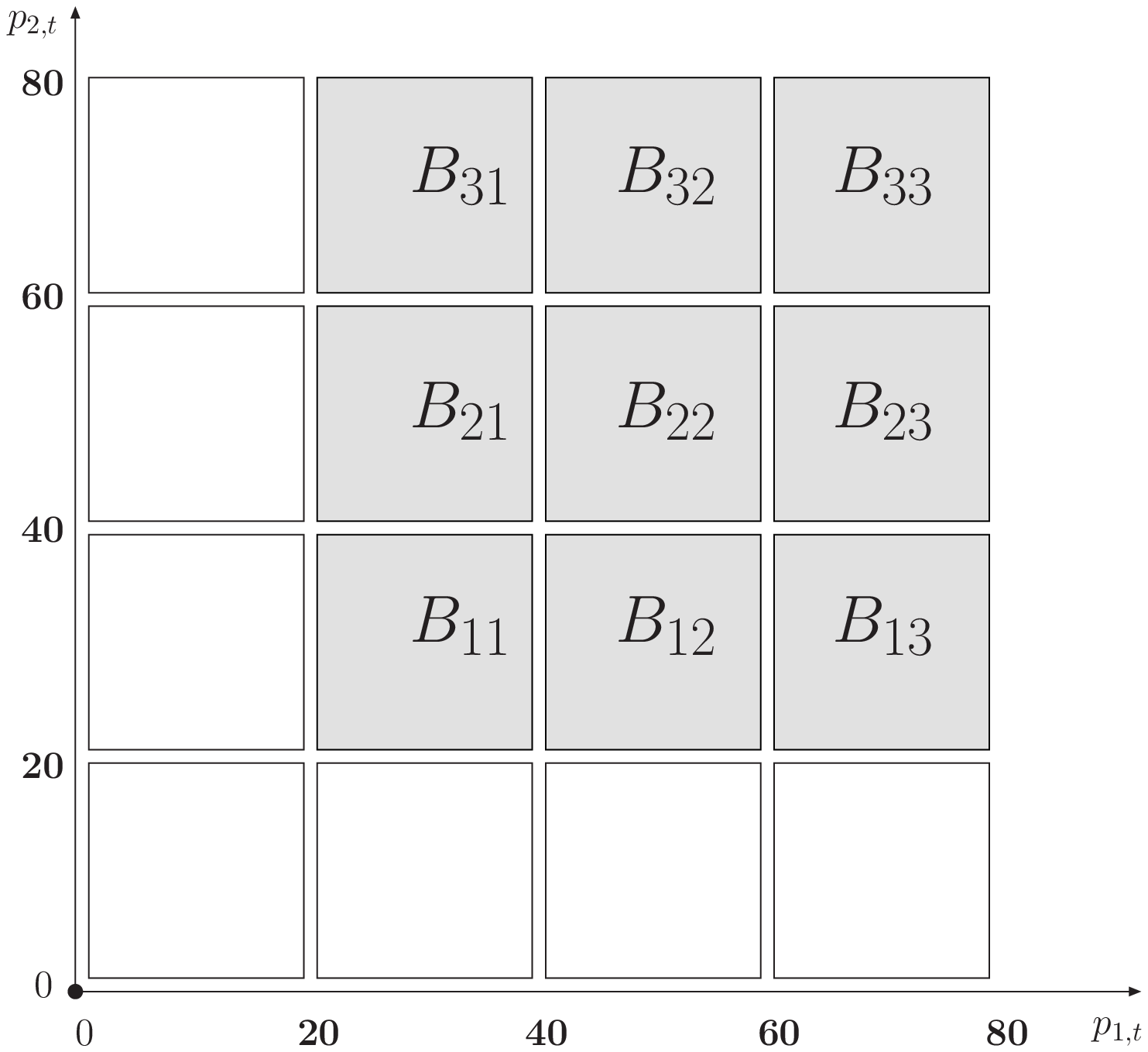}}
\caption{
The definition of the windows in
$(p_{1,t}, p_{2,t})$ plane for RHIC energy $\sqrt s = 200$~GeV (a) 
and for Tevatron energy $\sqrt s = 1960$~GeV (b).
\label{fig:ptpt_planes}
}
\end{center}
\end{figure}
%-----------

In Fig.\ref{fig:phid_200} we show angular azimuthal correlations
$d\sigma / d\phi_-$ at RHIC energy $\sqrt s = 200$~GeV for the 
Kwieci\'nski UPDFs
in the $k_t$-factorization approach with on-shell and off-shell matrix
elements and for the NLO collinear-factorization approach with extra
leading jet condition $p_{3,t} < p_{2,t}$. Here transverse momentum of
the photon and that of the associated jet $p_{1,t}, p_{2,t}$ belong
to the interval $(5, 20)$~GeV. There is almost no difference between
results obtained with off-shell and on-shell (see Appendix A) matrix
elements.

%------------------------------
\begin{figure}[!htb] % Figure 10
\begin{center}
\includegraphics[width=0.49\textwidth]{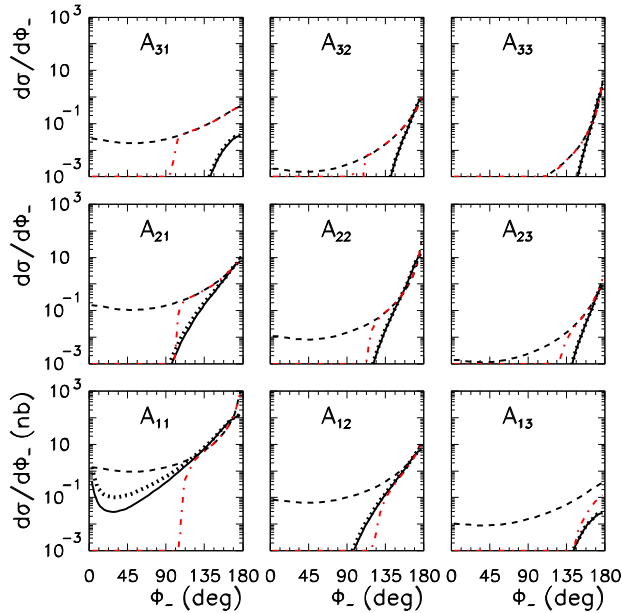}
\caption{
Angular azimuthal correlations $d\sigma / d\phi_-$ at $\sqrt s = 200$~GeV for
Kwieci\'nski on-shell ME (solid), Kwieci\'nski off-shell ME (thick dotted),
NLO collinear with no cuts on $p_{3,t}$ (dashed) and 
NLO collinear with cut on $p_{3,t} < p_{2,t}$ (dash-dotted).
Here $y_1, y_2 \in (-5,5)$. 
\label{fig:phid_200}
}
\end{center}
\end{figure}
%-----------

In Fig.\ref{fig:phid_1960} we show analogous angular distributions
as in Fig.\ref{fig:phid_200} 
but for Tevatron energy $\sqrt s = 1960$~GeV. 
In Fig.\ref{fig:phid_1960_abs_y_09} we show angular correlations
for a restricted range of rapidities
$|y_1|, |y_2| < 0.9$ (corresponding to the present Tevatron apparatus)
of the photon and the correlated jet. Limiting to midrapidities
does not change the shape of azimuthal correlations significantly.

%------------------------------
\begin{figure}[!htb] % Figure 11
\begin{center}
\includegraphics[width=0.49\textwidth]{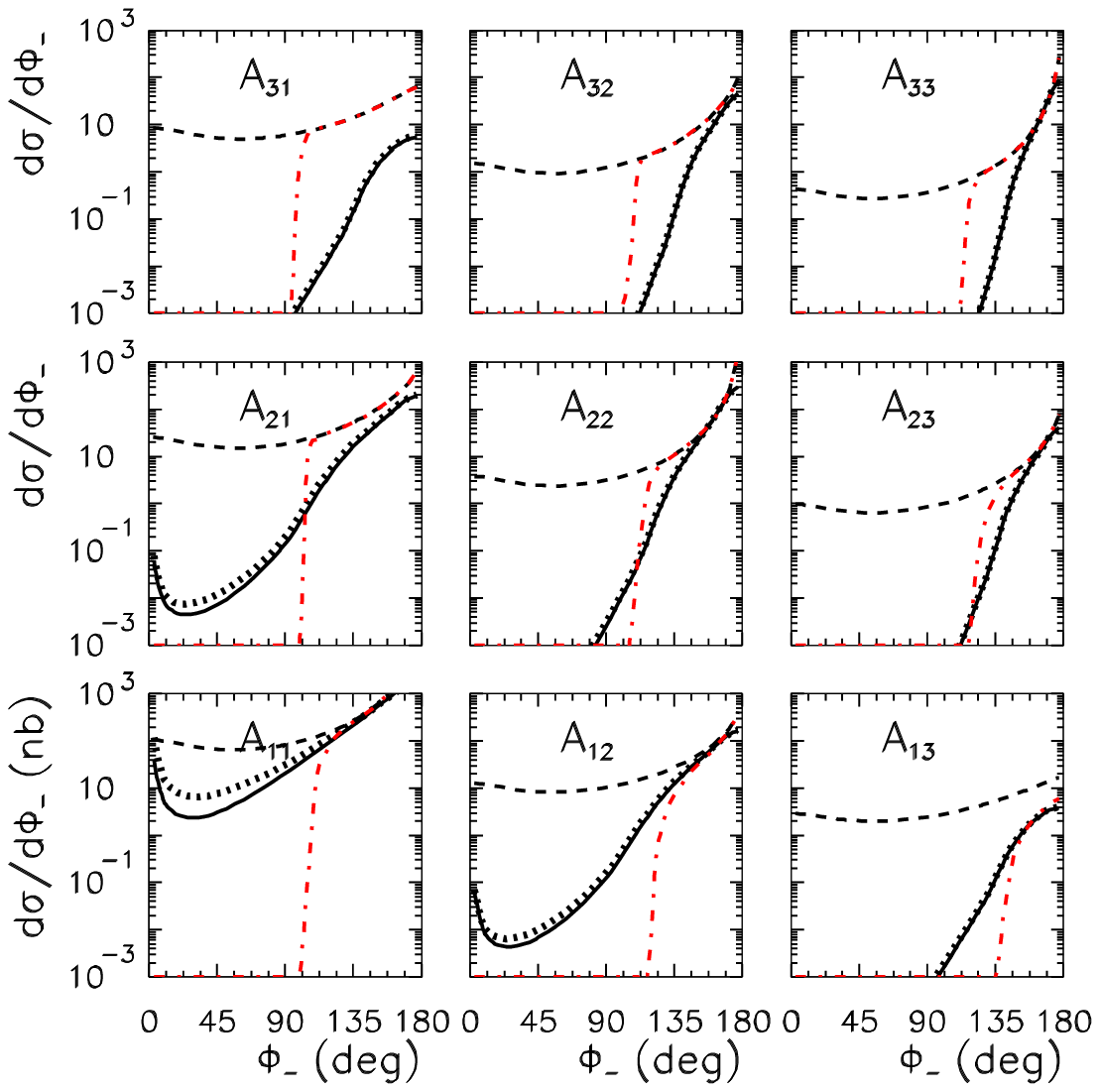}
\caption{
Angular azimuthal correlations $d\sigma / d\phi_-$ at $\sqrt s = 1960$~GeV for 
Kwieci\'nski on-shell ME (solid), Kwieci\'nski off-shell ME (thick dotted),
NLO collinear with no cuts on $p_{3,t}$ (dashed) and  
NLO collinear with cut on $p_{3,t} < p_{2,t}$ (dash-dotted).
Here $y_1, y_2 \in (-5,5)$. 
\label{fig:phid_1960}
}
\end{center}
\end{figure}
%-----------

%-------------------------------
\begin{figure}[!htb] % Figure 12
\begin{center}
\includegraphics[width=0.49\textwidth]{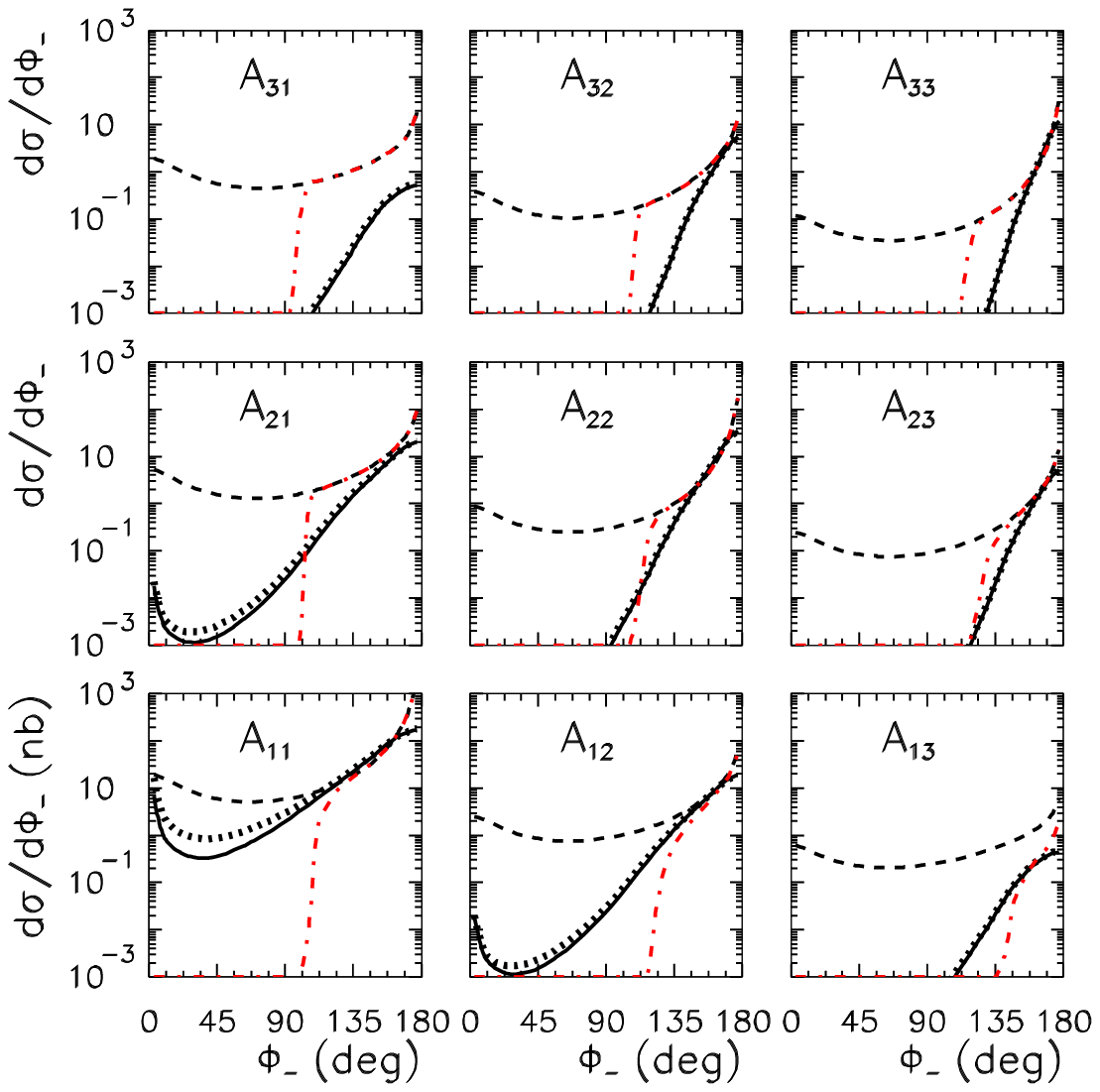}
\caption{
Angular azimuthal correlations $d\sigma / d\phi_-$ at $\sqrt s = 1960$~GeV for 
Kwieci\'nski on-shell ME (solid), Kwieci\'nski off-shell ME (thick dotted),
NLO collinear with no cuts on $p_{3,t}$ (dashed) and
NLO collinear with cut on $p_{3,t} < p_{2,t}$ (dash-dotted).
Here $|y_1|, |y_2| < 0.9$. 
\label{fig:phid_1960_abs_y_09}
}
\end{center}
\end{figure}
%-----------

In Fig.\ref{fig:phid_1960_high_pt} we show similar distributions 
as in Fig.\ref{fig:phid_1960} but for transverse-momentum windows
spanned over broader range of
transverse momenta $p_{1,t}, p_{2,t} \in (20, 80)$~GeV for the photon
and the jet.
We observe slightly faster decrease of the $k_t$-factorization
cross sections for larger $p_{1,t}$ and $p_{2,t}$.

% and/or when increasing
%azimuthal distance from the back-to-back configuration.

%-------------------------------
\begin{figure}[!htp] % Figure 13
\begin{center}
\includegraphics[width=.49\textwidth]{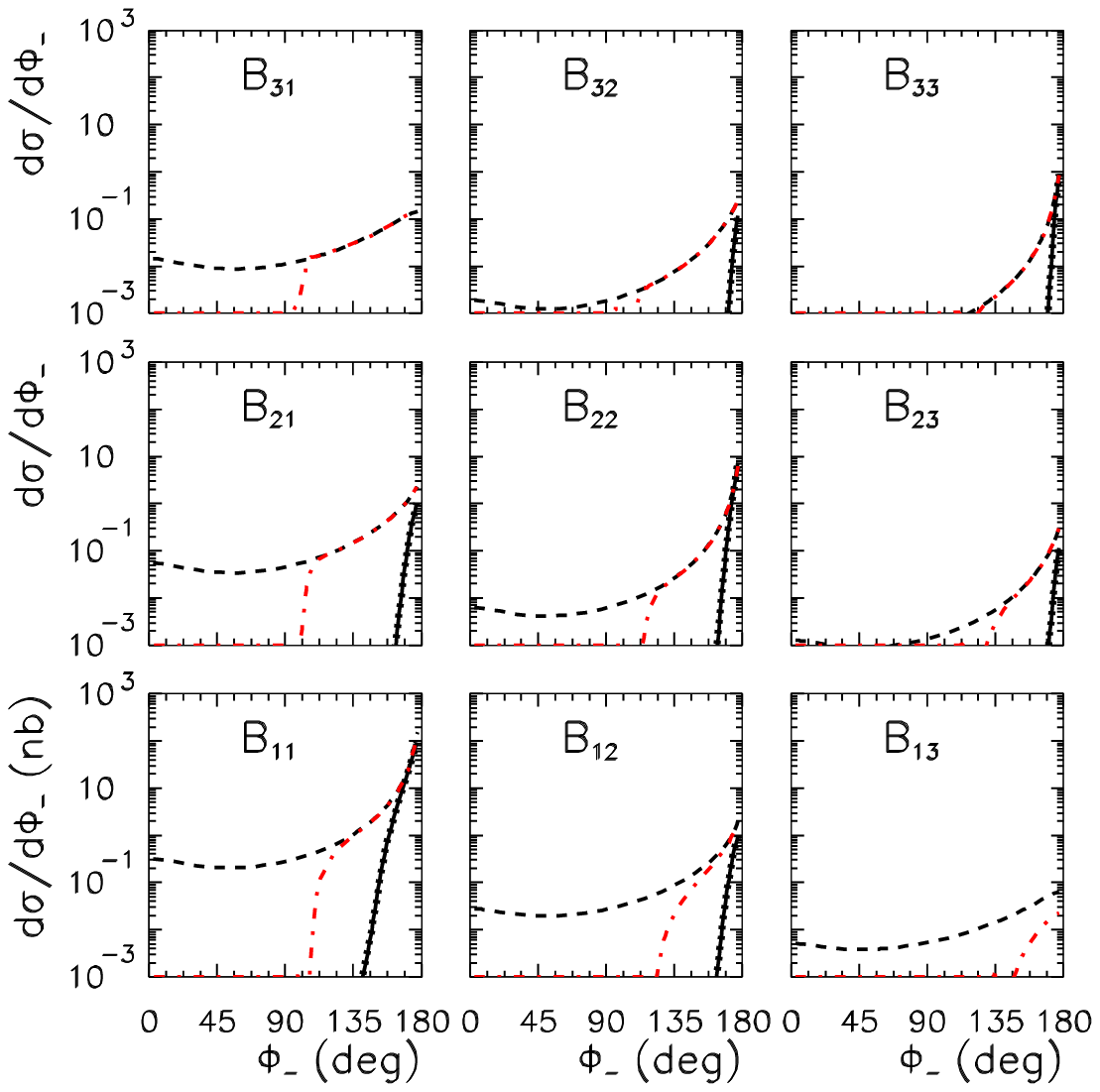}
\caption{
Angular azimuthal correlations $d\sigma / d\phi_-$ at $\sqrt s = 1960$~GeV for 
Kwieci\'nski on-shell ME (solid), Kwieci\'nski off-shell ME (thick dotted)
NLO collinear with no cuts on $p_{3,t}$ (dashed) and, 
NLO collinear with cut on $p_{3,t} < p_{2,t}$ (dash-dotted).
Here $y_1, y_2 \in (-5,5)$. 
\label{fig:phid_1960_high_pt}
}
\end{center}
\end{figure}
%-----------

The standard collinear approach can be applied only in the region
which is free of singularities. In order to eliminate the regions
where the pQCD calculation is not reliable some cuts on the measured
transverse momenta must be applied.
The simplest method is to use cuts shown in
Fig.\ref{fig:excluded_region}.
Mathematically this means that $p_{1,t} > p_{cut}$, $p_{2,t} > p_{cut}$
and 
\begin{eqnarray}
|p_{1,t} - p_{2,t}| > \Delta_S \; .
\label{delta_s}
\end{eqnarray}

%-------------------------------
\begin{figure}[!htb] % Figure 14
\begin{center}
\subfigure[]{\label{}
\includegraphics[height=6cm]{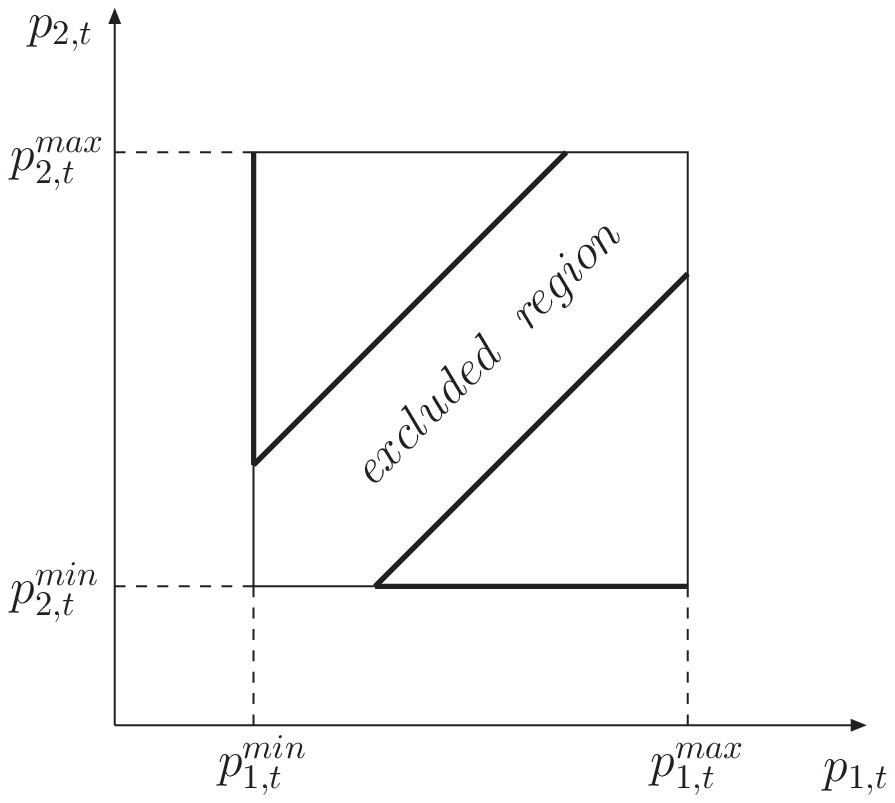}}
\caption{
Diagram showing excluded region in $(p_{1,t}, p_{2,t})$ plane.
\label{fig:excluded_region}
}
\end{center}
\end{figure}
%-----------
%
We shall call the last cut a scalar cut for further easy reference.
In Fig.\ref{fig:delta_s} we show azimuthal angle correlation function
for different values of the scalar cut $\Delta_S = 0,1,2,3$~GeV.
Clearly the NLO singularity at $\phi_{-} = \pi$ can be removed
by imposing the cut. However, the cut lowers also the
$k_t$-factorization cross section. 

%-------------------------------
\begin{figure}[!htp] % Figure 15
\begin{center}
\includegraphics[width=.49\textwidth]{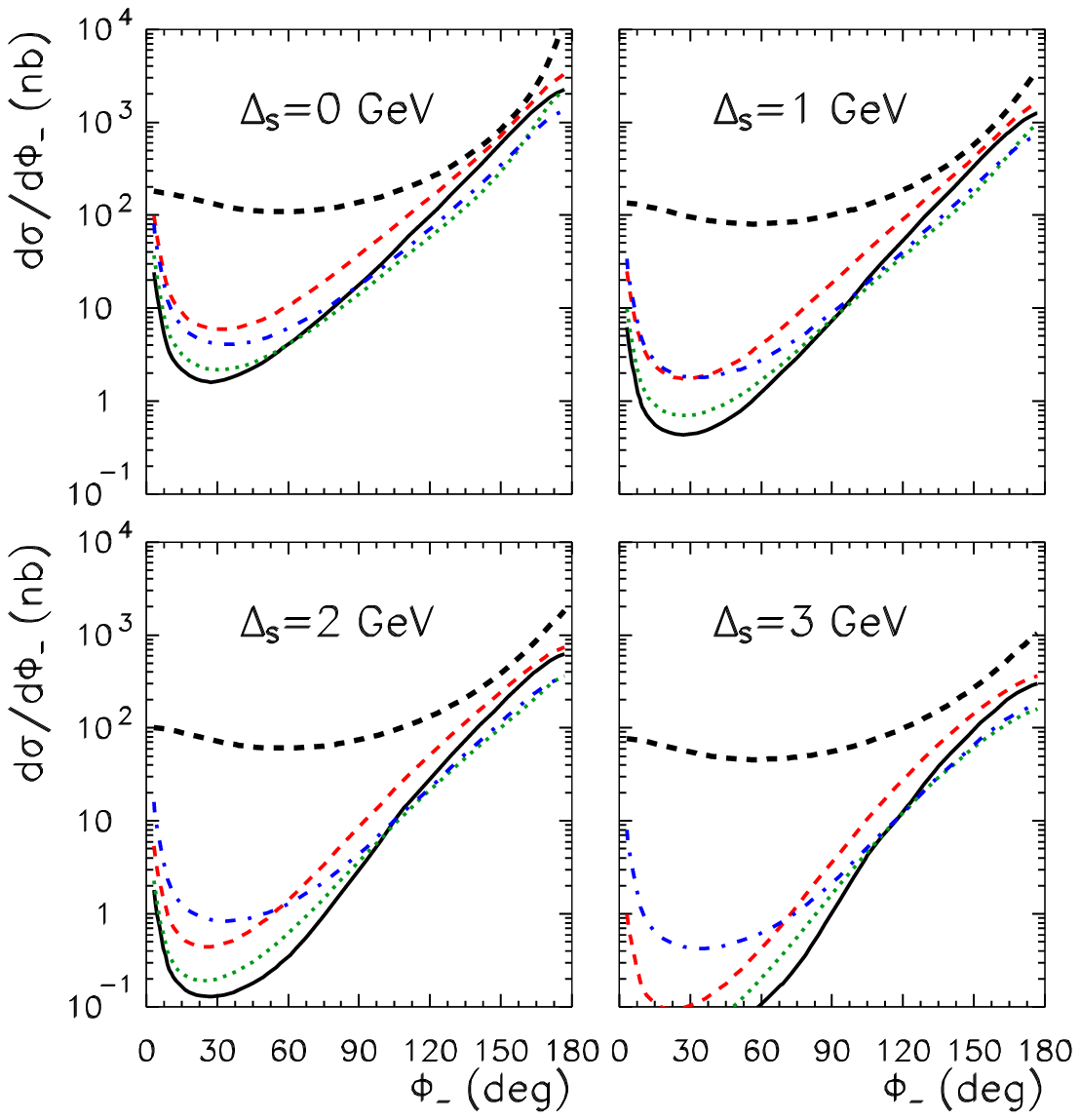}
\caption{
Angular azimuthal correlations $d\sigma / d\phi_-$ at $\sqrt s = 1960$~GeV for different 
(scalar) cuts $\Delta_S = 0,1,2,3$~GeV for NLO collinear (dashed),
Kwieci\'nski (solid), BFKL (dashed), KL (dotted) and 
KMR (dash-dotted). Here $p_{1,t},p_{2,t}\in (5,20)$~GeV 
and $y_1, y_2 \in (-5,5)$. 
\label{fig:delta_s}
}
\end{center}
\end{figure}
%-------------------------------

We have also tried another option to cut off the singularity:
\begin{eqnarray}
|\vec{p}_{1,t} + \vec{p}_{2,t}| > \Delta_V   \; .
\label{delta_v}
\end{eqnarray}
This type of the cut will be called vector one for brevity.
In Fig.\ref{fig:delta_v} we show corresponding photon-jet azimuthal
angle correlation function with different values of the cut 
$\Delta_V = 0,1,2,3$~GeV.
The situation here is very similar to that for the scalar cut.

%-------------------------------
\begin{figure}[!htp] % Figure 16
\begin{center}
\includegraphics[width=.49\textwidth]{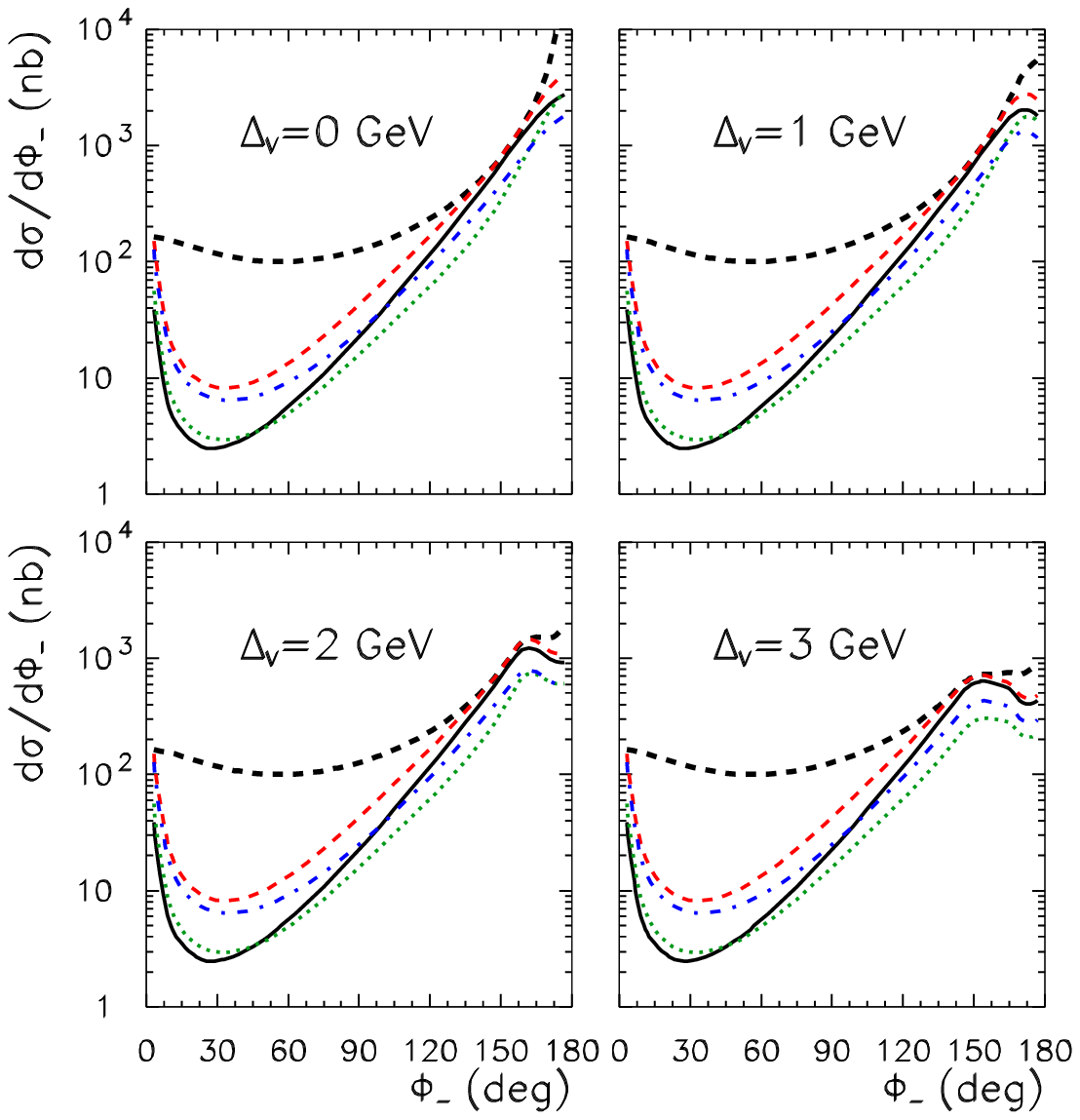}
\caption{
Angular azimuthal correlations $d\sigma / d\phi_-$ at $\sqrt s = 1960$~GeV for different 
(vector) cuts $\Delta_V = 0,1,2,3$~GeV for NLO collinear (dashed),
Kwieci\'nski (solid), BFKL (dashed), KL (dotted) and 
KMR (dash-dotted). Here $p_{1,t},p_{2,t}\in (5,20)$~GeV and 
$y_1, y_2 \in (-5,5)$. 
\label{fig:delta_v}
}
\end{center}
\end{figure}
%---------------------------------

\end{widetext}

%==============================
\section{Conclusions}
%==============================

We have performed for the first time the lacking in the literature
calculation of the photon-jet correlation observables in proton-proton
(RHIC) and proton-antiproton (Tevatron) collisions.
Up to now such correlations have not been studied experimentally either.
We have concentrated on the
region of small transverse momenta (semi-hard region) where
the $k_t$-factorization approach seems to be the most efficient and
theoretically justified tool.
We have calculated correlation observables for different unintegrated parton
distributions from the literature. Our previous analysis of
inclusive spectra of direct photons suggests that the Kwieci\'nski
distributions give the best description at low and intermediate
energies.
We have discussed the role of the evolution scale of the Kwieci\'nski
UPDFs on the azimuthal correlations. In general, the bigger the scale
the bigger decorrelation in azimuth is observed. When the scale
$\mu^2 \sim p_t^2$(photon) $\sim p_t^2$(associated jet)
(for the  kinematics chosen $\mu^2 \sim$ 100 GeV$^2$) is assumed, much bigger 
decorrelations can be observed than from the standard Gaussian smearing
prescription often used in phenomenological studies.

The correlation function depends strongly on whether it is the
correlation of the
photon and any jet or the correlation of the photon and the leading-jet
which is considered.
In the last case there are regions in azimuth and/or in the two-dimensional
($p_{1,t}, p_{2,t}$) space which cannot be populated in the standard
next-to-leading order approach. In the latter case the $k_t$-factorization
seems to be a useful and efficient tool.

We believe that the photon-jet correlations can be measured
at Tevatron. At RHIC one can measure jet-hadron correlations
for rather not too high transverse momenta of the trigger photon and of
the associated hadron. This is precisely the semihard region discussed here.
In this case the theoretical calculations would require inclusion of the
fragmentation process. This can be done easily assuming independent parton
fragmentation method using fragmentation functions extracted
from $e^+ e^-$ collisions. This will be a subject of the following analysis.

\vskip 0.5cm

{\bf Acknowledgments}
We are indebted to Jan Rak from the PHENIX collaboration for the
discussion of recent results for photon-hadron correlations at RHIC.
This work was partially supported by the grant
of the Polish Ministry of Scientific Research and Information Technology
number 1 P03B 028 28.

%-------------------------
\section{Appendix}
%-------------------------

%----------------------------------------------------------------------------------
\subsection{Matrix elements for $2 \to 2$ processes with initial off-shell partons}
%----------------------------------------------------------------------------------

In this paper we include four $2 \to 2$ processes such as 
$q\bar q \to \gamma g, \bar qq \to \gamma g, gq \to \gamma q, qg \to \gamma q$ 
important at midrapidity and relatively small transverse momenta.
The corresponding matrix elements for the on-shell initial partons read
\begin{eqnarray*}
\overline{|\mathcal{M}_{q \bar q \to \gamma g}|^2} 
&=&
\pi \alpha_{em}\sqrt{\alpha_{1,s}\alpha_{2,s}}
(16 \pi) \left(\frac{8}{9}\right)
\left(\frac{\hat u}{\hat t} + \frac{\hat t}{\hat u}\right)
\; , \\
\overline{|\mathcal{M}_{\bar q q\to \gamma g}|^2}
&=&
\pi \alpha_{em}\sqrt{\alpha_{1,s}\alpha_{2,s}}
(16 \pi) \left(\frac{8}{9}\right)
\left(\frac{\hat t}{\hat u} + \frac{\hat u}{\hat t}\right)
\; , \\
\overline{|\mathcal{M}_{g q\to \gamma q}|^2}
&=&
\pi \alpha_{em}\sqrt{\alpha_{1,s}\alpha_{2,s}}
(16 \pi) \left(-\frac{1}{3}\right)
\left(\frac{\hat u}{\hat s} + \frac{\hat s}{\hat u}\right)
\; , \\
\overline{|\mathcal{M}_{q g\to \gamma q}|^2}
&=&
\pi \alpha_{em}\sqrt{\alpha_{1,s}\alpha_{2,s}}
(16 \pi) \left(-\frac{1}{3}\right)
\left(\frac{\hat t}{\hat s} + \frac{\hat s}{\hat t}\right)
\; . 
\label{on_shell_matrix_elements}
\end{eqnarray*}
The matrix elements for the off-shell initial partons were derived
in Ref.\cite{LZ_photon}.
To a good approximation the matrix elements for the off-shell initial
partons can be also obtained by using the on-shell formulae
(\ref{on_shell_matrix_elements}) but with
$\hat s, \hat t, \hat u$ calculated including off-shell initial
kinematics.
In this case $\hat s+ \hat t+ \hat u = k_1^2 + k_2^2$, where
$k_1^2, k_2^2 < 0$ denote virtualities of initial partons.
Our prescription can be treated as a smooth analytic continuation 
of the on-shell formula off mass shell. With our choice of initial
parton four-momenta $k_1^2 = -k_{1,t}^2$ and $k_2^2 = -k_{2,t}^2$.

Explicit formulae for exact off-shell matrix elements were calculated
and can be found in Ref.\cite{LZ_photon}. In this paper we compare results
obtained with both (approximate and exact) ways.

%---------------------------------------------------
\subsection{Matrix elements for $2 \to 3$ processes}
%---------------------------------------------------

In order to obtain parton-parton $\to$ $\gamma$-jet matrix elements 
for the next-to-leading order one can use the following  
expression 
%
%\begin{eqnarray*}
\begin{equation*}
\begin{split}
\frac{1}{4}\sum_{spins}&\frac{1}{N_C}\sum_{col}|M|^2 
=
C_F4\pi\alpha e_q^2 g_{1,s}^2 g_{2,s}^2\\
&\times \left[2(C_F-\frac{1}{2}N_C)a_4 + N_C
\frac{a_2a_7 + a_3a_6}{a_9} \right] \nonumber \\ 
&\times
\left[\frac{a_1^2 + a_5^2}{a_2a_3a_6a_7}
    + \frac{a_2^2 + a_6^2}{a_1a_3a_5a_7}
    + \frac{a_1^3 + a_7^2}{a_1a_2a_5a_6} \right]
\end{split}
\end{equation*}
%\end{eqnarray*}
%
for the $\gamma(p_1)+q(p_2) \to g(k_1)+g(k_3)+q(k_2)$ process obtained in \cite{Aurenche87}.
Here $C_F = 4/3$, $N_C = 3$, where
\begin{eqnarray*}
g_{1,s}^2 &=& 4\pi \alpha_s(p_{1,t}^2) \\
g_{2,s}^2 &=& 4\pi \alpha_s(p_{2,t}^2)
\end{eqnarray*}
and
\begin{eqnarray*}
\begin{matrix}
a_1=p_2 \cdot p_1, & a_5=k_2 \cdot p_1, & a_8=k_3 \cdot p_1, & a_{10}=k_1 \cdot p_1, \\
a_2=p_2 \cdot k_1, & a_6=k_2 \cdot k_1, & a_9=k_3 \cdot k_1, \\
a_3=p_2 \cdot k_3, & a_7=k_2 \cdot k_3, \\
a_4=p_2 \cdot k_2, \\
\end{matrix}
\end{eqnarray*}
are redundant invariants.
% and also
%
%\begin{eqnarray*}
%\begin{matrix}
%p_i = (p_i^0,\; p_i^x,\; p_i^y,\; p_i^z)& i = 1, 2 \\
%k_i = (k_i^0,\; k_i^x,\; k_i^y,\; k_i^z)& i = 1, 2, 3 \\
%\end{matrix}
%\end{eqnarray*}
%
%where
%
%\begin{eqnarray*}
%p_1 &=& \left(x_1 \frac{\sqrt s}{2},\; 0,\; 0,\; x_1 \frac{\sqrt s}{2}\right) \\ 
%p_2 &=& \left(x_2 \frac{\sqrt s}{2},\; 0,\; 0,\; -x_2 \frac{\sqrt s}{2}\right) \\ 
%k_1 &=& \left(p_{1,t} \cosh (y_1),\; p_{1,t}^x,\; p_{1,t}^y,\; p_{1,t} \sinh (y_1)\right) \\
%k_2 &=& \left(p_{2,t} \cosh (y_2),\; p_{2,t}^x,\; p_{2,t}^y,\; p_{2,t} \sinh (y_2)\right) \\
%k_3 &=& \left(p_{3,t} \cosh (y_3),\; p_{3,t}^x,\; p_{3,t}^y,\; p_{3,t} \sinh (y_3)\right) 
%\end{eqnarray*}
%
The longitudinal momentum fractions are calculated as:
\begin{eqnarray*}
x_1 &=& (p_{1,t}e^{\phantom{-}y_1}+p_{2,t}e^{\phantom{-}y_2}+p_{3,t}e^{\phantom{-}y_3})/\sqrt{s}\\
x_2 &=& (p_{1,t}e^{ -y_1}+p_{2,t}e^{ -y_2}+p_{3,t}e^{ -y_3})/\sqrt{s}
\end{eqnarray*}
%
%and
%
%\begin{eqnarray*}
%p_{1,t}^x &=& p_{1,t}\cos(\phi_1) \\
%p_{1,t}^y &=& p_{1,t}\sin(\phi_1) \\ \\
%p_{2,t}^x &=& p_{2,t}\cos(\phi_2) \\
%p_{2,t}^y &=& p_{2,t}\sin(\phi_2) \\ \\
%p_{3,t}^x &=& - p_{1,t}^x - p_{2,t}^x  \\
%p_{3,t}^y &=& - p_{1,t}^y - p_{2,t}^y 
%\end{eqnarray*}
%
%where
%
%\begin{eqnarray*}
%\phi_1 &=& (\phi_+ + \phi_-)/2 \\
%\phi_2 &=& (\phi_+ - \phi_-)/2
%\end{eqnarray*}
%
%and here arbitrary phase $\phi_+ = 0$.

\begin{figure}[!htb] % Figure 17
\begin{center}
\includegraphics[height=2cm]{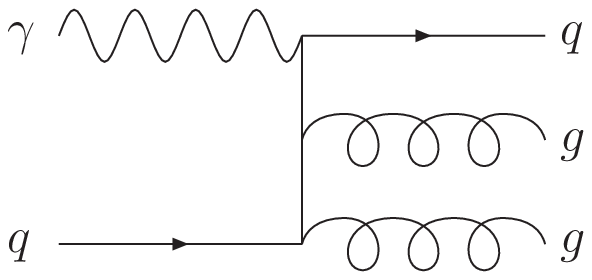} 
\caption{
Diagram of the $\gamma q \to g g q$ process.
\label{fig:gamq_diagram}
}
\end{center}
\end{figure}

As an example the expression for matrix elements for the second diagram
\[
g(p_1)+g(p_2) \to \gamma(k_1)+q(k_3)+ \bar q(k_2) 
\]
in Fig.\ref{nlo_b} we get from
\[
\overbrace{\gamma(p_1)}+\underbrace{q(p_2)} \to 
\overbrace{g(k_1)}+\underbrace{g(k_3)} + q(k_2) 
\]
diagram (see Fig.\ref{fig:gamq_diagram}) if we make the following replacement
\begin{eqnarray*}
\begin{matrix}
p_1 \to k_1, \\
k_1 \to p_1, \\
p_2 \to k_3, \\
k_3 \to p_2, \\
\end{matrix}
\end{eqnarray*}

%---------------------
%---------------------

thus obtaining:
\begin{eqnarray*}
\begin{matrix}
a_1 \to a_9, & a_5 \to a_6, & a_8 \to a_2, & a_{10} \to a_{10}, \\
a_2 \to a_8, & a_6 \to a_5, & a_9 \to a_1, \\
a_3 \to a_3, & a_7 \to a_4, \\
a_4 \to a_7, \\
\end{matrix}
\end{eqnarray*}
%
%%-----------------
%\subsection{Running $\alpha_s$}
%%-----------------
% 
%The treatment of the running coupling constants in $2 \to 2$ and $2 \to 3$ subprocesses
%is quite important in numerical evaluation of the cross section.
%For the $2 \to 2$ case we shall try several prescriptions:
%%
%\begin{eqnarray*}
%(a)\;\;\;\; \alpha_s^2 &=& \alpha_s(p_{1,t}^2)
%(b)\;\;\;\; \alpha_s^2 &=& \alpha_s(p_{2,t}^2)
%(c)\;\;\;\; \alpha_s^2 &=& \alpha_s(p_{1,t}p_{2,t})
%\label{}
%\end{eqnarray*}
%

\newpage
%==========================

%====================

\begin{thebibliography}{99}

\bibitem{D0_dijets}
S.S. Adler et al. (PHENIX collaboration), Phys. Rev. Lett. {\bf 97}
(2006) 052301; \\
S.S. Adler et al. (PHENIX collaboration), Phys. Rev. {\bf C73} (2006)
054903;\\
S.S. Adler et al. (PHENIX collaboration), Phys. Rev. Lett. {\bf 96}
(2006) 222301;\\
M. Oldenburg et al. (STAR collaboration), Nucl. Phys. {\bf A774} (2006) 507.

\bibitem{RHIC_hadron_hadron}
S.S. Adler et al. (PHENIX collaborations), Phys. Rev. {\bf D74} (2006) 
072002. 

\bibitem{RHIC_photon_hadron}
DongJo Kim, a talk at the international workshop on ``High-$p_t$ processes
at LHC'', Jyv\"askyl\"a, Finland, March 23-28, 2007.

\bibitem{Berends}
F.A. Berends, R. Kleiss, P.De Causmaecker, R. Gastmans and T.T. Wu, 
Phys. Lett. {\bf B 103} 124 (1981).

\bibitem{Aurenche87}
P.~Aurenche, A.~Baier, A.~Douiri, M.~Fontannaz and D.~Schiff, 
Nucl. Phys. {\bf B286} 553 (1987).

\bibitem{PS06_photon}
T.~Pietrycki and A.~Szczurek, Phys. Rev. {\bf D75} 014023 (2007). 

\bibitem{LZ_photon}
A.V. Lipatov and N.P. Zotov,
Phys. Rev. {\bf D72} 054002 (2005);\\
A.V. Lipatov and N.P. Zotov, hep-ph/0507243.

\bibitem{Mariotto}
  C.~B.~Mariotto, M.~B.~Gay Ducati and M.~V.~T.~Machado,
  %``Heavy quark photoproduction in k(T) factorization approach,''
  Phys.\ Rev.\ {\bf D 66} 114013 (2002)
  [arXiv:hep-ph/0208155].

\bibitem{LS04}
M.~{\L}uszczak and A.~Szczurek, Phys.\ Lett.\ {\bf B 594} 291 (2004).

\bibitem{BS00}
S.P. Baranov and M. Smizanska, Phys. Rev. {\bf D62} 014012 (2000).

\bibitem{LS06}
M.~{\L}uszczak and A.~Szczurek, arXiv:hep-ph/0512120, 
Phys. Rev. {\bf D73} 054028 (2006).

\bibitem{HKSST1}
  P.~Hagler, R.~Kirschner, A.~Schafer, L.~Szymanowski and O.~V.~Teryaev,
  %``Direct J/psi hadroproduction in k(T)-factorization and the color octet
  %mechanism,''
  Phys.\ Rev.\ {\bf D 63} 077501 (2001)
  [arXiv:hep-ph/0008316].

\bibitem{HKSST2}
  P.~Hagler, R.~Kirschner, A.~Schafer, L.~Szymanowski and O.~V.~Teryaev,
  %``Towards a solution of the charmonium production controversy: k(T)
  %factorization versus color octet mechanism,''
  Phys.\ Rev.\ Lett.\  {\bf 86} 1446 (2001)
  [arXiv:hep-ph/0004263].

\bibitem{KS04}
J. Kwieci\'nski and A. Szczurek, Nucl. Phys. {\bf B680} 164 (2004).

\bibitem{LZ05}
  A.~V.~Lipatov and N.~P.~Zotov,
  %``Higgs boson production at hadron colliders in the k(T)-factorization
  %approach,''
  Eur.\ Phys.\ J.\ {\bf C 44} 559 (2005)
  [arXiv:hep-ph/0501172];\\
    A.~V.~Lipatov and N.~P.~Zotov,
  %``Higgs production via gluon fusion with k(T) factorization,''
  arXiv:hep-ph/0510043.

\bibitem{LS05}
M. {\L}uszczak and A. Szczurek, hep-ph/0504119,
Eur. Phys. J. {\bf C46} 123 (2006).

\bibitem{Kwiecinski}
J. Kwieci\'nski, Acta Phys. Polon. {\bf B33} 1809 (2002);\\
A. Gawron and J. Kwieci\'nski, Acta Phys. Polon. {\bf B34} 133
(2003);\\
A. Gawron, J. Kwieci\'nski and W. Broniowski, Phys. Rev. {\bf D68} 054001
(2003).

\bibitem{KMR}
M.A. Kimber, A.D. Martin and M.G. Ryskin,
Phys. Rev. {\bf D63} 114027 (2001).

\bibitem{Owens}
J.F. Owens,
Rev. Mod. Phys. {\bf 59} 465 (1987).

\bibitem{AM04}
U. d'Alesio and F. Murgia,
Phys. Rev. {\bf D70} 074009 (2004).

\bibitem{GRV98}
M. Gl\"uck, E. Reya and A. Vogt, Eur. Phys. J. {\bf C5} 461 (1998).


\end{thebibliography}
\end{document}